 \documentclass[journal]{IEEEtran}
\usepackage{epsfig}
\usepackage{graphicx}
\usepackage{amsmath}
\usepackage{amssymb}
\usepackage{amsmath} 
\usepackage{amsmath} 
\usepackage{algorithmic}
\usepackage{algorithm}
\usepackage{subfigure}
\usepackage{multirow}
\usepackage{url}
\usepackage{color}
\usepackage{hyperref}
\usepackage{balance}
\usepackage{setspace}
\usepackage{slashbox}
\usepackage{booktabs}
\usepackage{threeparttable}
\renewcommand{\arraystretch}{1.5} 
\renewcommand{\baselinestretch}{0.97}

\newcommand{\ie}{{\em i.e.}}           

\ifCLASSOPTIONcompsoc
  \usepackage[nocompress]{cite}
\else
  \usepackage{cite}
\fi

\ifCLASSINFOpdf
\else
\fi

\begin{document}
%
\title{Crossover-Net: Leveraging the Vertical-Horizontal Crossover Relation for Robust Segmentation}
%
%
%
%
\author{Qian~Yu, Yinghuan~Shi, Yefeng Zheng, Yang~Gao, Jianbing~Zhu, Yakang~Dai
\IEEEcompsocitemizethanks{\IEEEcompsocthanksitem Qian Yu, Yinghuan Shi and Yang Gao are with the State Key Laboratory for Novel Software Technology, Nanjing University, China. Qian Yu is also with School of Data and Computer Science, Shandong Women's University, China (e-mail: yuqian@sdwu.edu.cn; syh@nju.edu.cn; gaoy@nju.edu.cn).
\IEEEcompsocthanksitem Yefeng Zheng is with Youtu Lab, Tencent, China (e-mail: yefengzheng@tencent.com).
\IEEEcompsocthanksitem Jianbing Zhu is with the Suzhou Science and Technology Town Hospital, China. (e-mail: zeno1839@126.com)
\IEEEcompsocthanksitem Yakang Dai is with the Suzhou Institute of Biomedical Engineering and Technology, Chinese Academy of Sciences, China. (e-mail: daiyk@sibet.ac.cn)%
 }
\thanks{}}

%
%


\IEEEtitleabstractindextext{%
\begin{abstract}
Robust segmentation for non-elongated tissues in medical images is hard to realize due to the large variation of the shape, size, and appearance of these tissues in different patients. In this paper, we present an end-to-end trainable deep segmentation model termed Crossover-Net for robust segmentation in medical images. Our proposed model is inspired by an insightful observation: during segmentation, the representation from the horizontal and vertical directions can provide different local appearance and orthogonality context information, which helps enhance the discrimination between different tissues by simultaneously learning from these two directions. Specifically, by converting the segmentation task to a pixel/voxel-wise prediction problem, firstly, we originally propose a cross-shaped patch, namely crossover-patch, which consists of a pair of (orthogonal and overlapped) vertical and horizontal patches, to capture the orthogonal vertical and horizontal relation. Then, we develop the Crossover-Net to learn the vertical-horizontal crossover relation captured by our crossover-patches. To achieve this goal, for learning the representation on a typical crossover-patch, we design a novel loss function to (1) impose the consistency on the overlap region of the vertical and horizontal patches and (2) preserve the diversity on their non-overlap regions. We have extensively evaluated our method on CT kidney tumor, MR cardiac, and X-ray breast mass segmentation tasks. Promising results are achieved according to our extensive evaluation and comparison with the state-of-the-art segmentation models.

\end{abstract}

}

\maketitle
\IEEEdisplaynontitleabstractindextext
\IEEEpeerreviewmaketitle
\bigskip
\bigskip

\IEEEraisesectionheading{\section{Introduction}}
\label{intro}

\IEEEPARstart{I}{n} medical images, \emph{e.g.}, computed tomography (CT), magnetic resonance imaging (MRI), and X-ray images, segmentation of non-elongated tissues is a significant yet challenging task. Basically, the representative non-elongated tissues contain the liver, heart, kidney, and brain, \emph{etc}. The purpose of segmenting these tissues is to provide the position, size, and intensity information to physicians for the subsequent diagnosis of liver cancer, cardiac disease, \emph{etc}. Unfortunately, due to the subjective (\emph{e.g.}, inaccurate delineation) and objective (\emph{e.g.}, massive images) factors, manual delineation is not desirable in clinical practice. Therefore, automatic segmentation methods for non-elongated tissues are in a great demand according to the aforementioned issues.
\begin{figure}[tb]
\centering
\setlength{\abovecaptionskip}{-0.1cm}   
\includegraphics[width =3.4in]{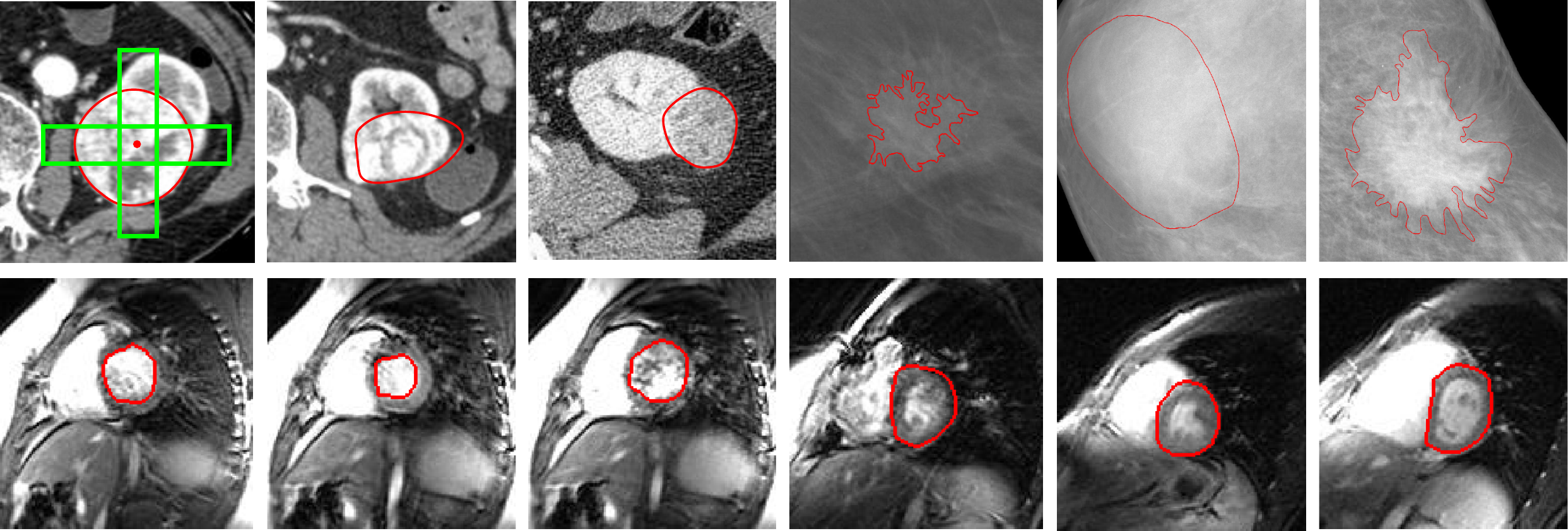}
\caption{Typical non-elongated tissues. The images in the first row are kidney tumors and breast masses, respectively. The second row shows some examples of endocardium and epicardium. The red contour indicates the ground truth of each tissue. The top-left image shows the example of our crossover-patch. }\label{tumors_examples}
\end{figure}
However, it is known that the segmentation of these tissues in an automatic manner is a challenging task. Taking kidney tumors in CT images as an example, as shown in Fig. \ref{tumors_examples}, the tumors could possibly appear anywhere in the kidney with different sizes, shapes and textures. Also, the intensity varies significantly even within the same tumor. In fact, most types of tumors also show these aforementioned characteristics as kidney tumors, such as X-ray breast masses (Fig. \ref{tumors_examples}). These characteristics pose a great challenge for segmentation methods. To clearly illustrate this observation, we also list some representative examples of MR cardiac images in Fig. \ref{tumors_examples}. It is obvious that the boundaries of the left ventricle endocardium and epicardium are fuzzy in a similar way.

Recently, many methods have been proposed to segment these non-elongated tissues, while these aforementioned challenges largely restrict the performance of the existing methods \cite{yang2018auto, zhu2018Adversarial, zhang2018photoacoustic, jiang2019multiple, khened2018densely, patravali20172d-3d}. Common models designed for general medical image segmentation (\emph{e.g.}, U-Net \cite{ronneberger2015u}, SegCaps \cite{lalonde2018capsules}) could not obtain very satisfactory results when they are directly applied to segment these tissues according to our following evaluation. To solve the above problems, we have developed Crossbar-Net \cite{yu2019crossbar} to segment kidney tumors in CT images, showing the promising results. However, the Crossbar-Net follows a conventional cascaded learning framework, where several sub-models are required to be trained round by round. Besides, the two orthogonal patches in Crossbar-Net are utilized separately and the information in the overlap region is thus ignored. As a step further, in this study, we wish to propose a more efficient and effective convolutional neural network (CNN) to segment non-elongated tissues. To achieve this goal, we intend to take into account the following two issues:
\begin{itemize}
 \item \textbf{End-to-end Training}. To reduce (1) the \emph{high computational burden} and (2) the \emph{invertible influence} from the different settings caused by the cascaded training, we wish to design a double-branch segmentation model to encode the vertical and horizontal information that could be trained in an end-to-end manner.
\item \textbf{Efficient Modeling}. Since separately sampling and utilizing the vertical and horizontal patches might waste the overlap region and bring in a large quantity of redundant patches, we try to learn a consistent representation between the vertical and horizontal patches if they have an overlap region, which could largely improve the utilization of information.
\end{itemize}

According to the above analysis, we propose an orthogonal patch pair in this paper, namely crossover-patch (see Fig. \ref{tumors_examples}). Basically, a crossover-patch consists of a vertical patch and a horizontal patch, with each patch fully covering the whole tissue along one direction (\emph{i.e.}, vertical or horizontal) from side-to-side. Based on the crossover-patch, we construct an end-to-end deep segmentation model, termed as Crossover-Net, to convert the segmentation task to a pixel/voxel-wise classification problem. Specifically, our Crossover-Net intends to classify the central pixel of a crossover-patch belonging to the target tissue or not. For a given pixel, the vertical and horizontal patches of its corresponding crossover-patch are utilized as a whole in Crossover-Net.
Overall, the technical contributions of this work can be summarized into the following four folds:

\begin{itemize}
\item For a given location (\emph{i.e.}, pixel or voxel) in the medical image, our proposed crossover-patch can not only capture the vertical and horizontal information separately but also provide the crossover relation information to Crossover-Net for the efficient and discriminative learning.
\item Our Crossover-Net is a dual-branch end-to-end segmentation model. Our model can learn the feature representation of the cross-shaped patches from two directions simultaneously. We also notice that if the feature representation in one direction is not discriminative enough, the other direction could benefit the current one from a complement perspective.
\item We design a novel loss function with the goal of making the feature representations learned from (1) the overlap region of each crossover-patch as consistent as possible, while (2) the non-overlap regions of the patch as different as possible. Thus, the target-guided information could be effectively highlighted.
\item Our model is easy to implement. According to our evaluation, our model can achieve the promising performance on various non-elongated tissue segmentation tasks.
\end{itemize}

\begin{figure*}[!htb]
\centering
\setlength{\abovecaptionskip}{-0.1cm}   
\setlength{\belowcaptionskip}{-2cm}   
 \includegraphics[width = 6.4in]{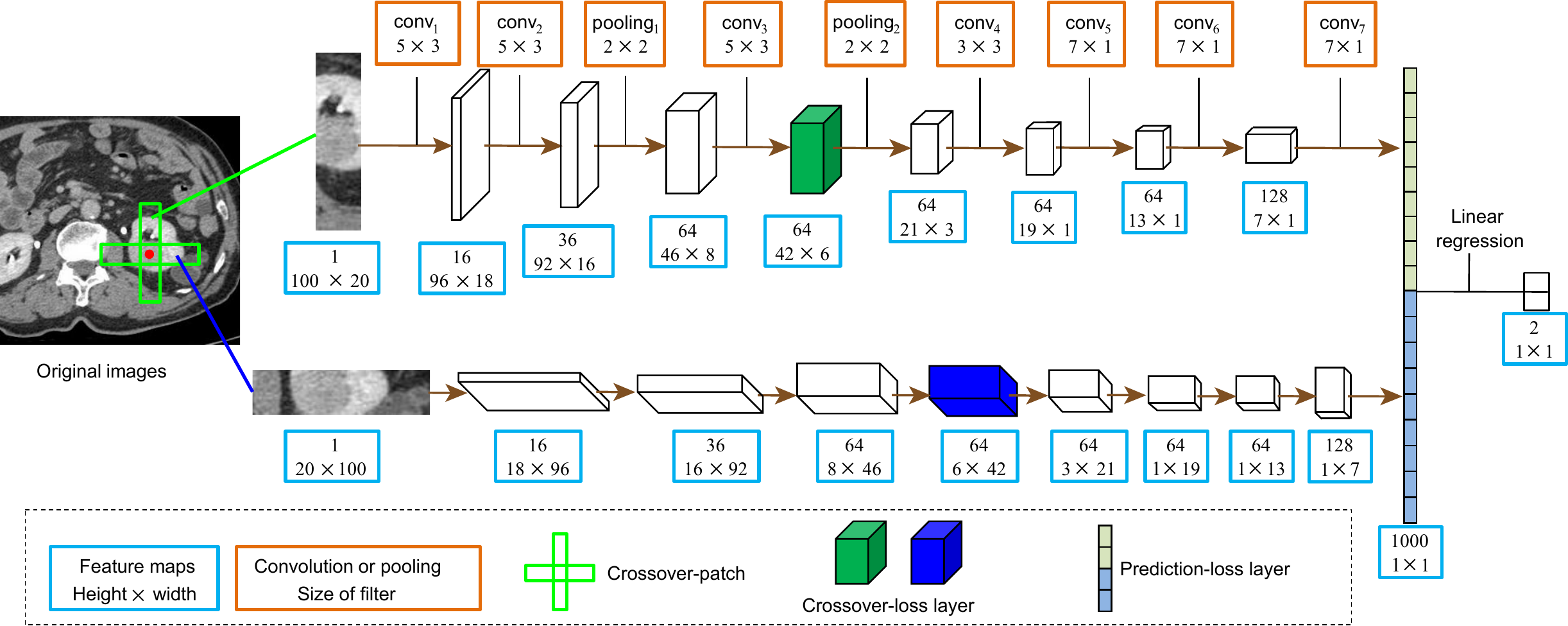}
\caption{Illustration of the architecture of our proposed Crossover-Net.}\label{framework}
\end{figure*}
\section{Related Work}
\label{sec:related-work}
Previous CNN-based segmentation methods can be roughly classified into two categories: the \emph{image-based} models and the \emph{patch-based} models. For the image-based models, the fully convolutional network (FCN) proposed in \cite{long2015fully} and its variants are widely used in various image segmentation tasks. Among all the FCN-style structure, for medical image segmentation, the U-Net \cite{ronneberger2015u} is a representative model. Currently, many state-of-the-art segmentation models are inspired by the merits of U-Net. For example, Lalonde \emph{et al.} \cite{lalonde2018capsules} designed a common model based on U-Net with capsules, SegCaps, which achieved promising results in many segmentation tasks. Similarly, the H-DenseUNet \cite{li2018h-denseunet:} is a hybrid U-Net model to fuse 2D and 3D features for liver tumor segmentation in CT images. For the patch-based model, Ciresan \emph{et al.} \cite{ciresan2012deep} employed multiple deep networks to segment biological neuron membranes by extracting the square patches in multi-scales using sliding-window. Wang \emph{et al.} \cite{wang2017central} devised a multi-branch CNN model to segment the lung nodules. Shi \emph{et al.} \cite{shi2017does} proposed a cascaded deep domain adaptation model to segment the prostate in CT images.

Recently, training deep models in a multi-scale or multi-branch manner to extract comprehensive features has aroused considerable interests. For example, Havaei \emph{et al.} \cite{havaei2017brain} and Razzak \emph{et al.} \cite{razzak2018effi} segmented the brain tumors with a two-pathway CNN architecture. Similarly, Moeskops \emph{et al.} \cite{moeskops2016automatic} presented a multi-scale approach to segmenting MR brain images. Also, Kamnitsas \emph{et al.} \cite{kamnitsas2017efficient} proposed a multi-scale 3D-CNN with two parallel pathways to segment brain lesions. In \cite{zhou2018high}, a multi-channel multi-scale CNN model was designed to reconstruct the plane-wave ultrasound images. Bien \emph{et al.} \cite{ng2018deep} also developed a multi-branch model, namely MRNet, to detect the general abnormalities and specific diagnoses on knee MR images. All these methods utilize the multi-scale input data to obtain high accuracy, and some other methods aggregate multi-scale features from a single input data to achieve good performance. For instance, Lin \emph{et al.} \cite{lin2018multi-scale} designed a multi-scale context intertwining strategy to aggregate features from different scales for pixel level semantic segmentation. Huang \emph{et al.} \cite{huang20183d} cropped multi-level features for colorectal tumor segmentation task. Also, methods in \cite{lin2019zigzagnet:, orsic2019in, li2019dfanet:, xiao2018unified, tian2019decoders} propagated and fused multi-scale features to construct different scale context for accurate segmentation.

From the perspective of the loss function, beyond traditional loss functions (\emph{e.g.}, the cross-entropy loss, the logistic loss, and the mean squared error loss), it is a new trend to design special loss functions to improve the performance of a deep segmentation model by using the prior knowledge of the segmentation task. For instance, in V-Net \cite{milletari2016v} (a typical 3D medical image segmentation model), a Dice similarity loss function was utilized to segment the MR prostate. Also, in 3D RU-Net \cite{huang20183d}, a hybrid Dice-based loss function was designed to help the model to locate the region of interesting (ROI) and segment the colorectal tumor simultaneously. In addition, a linear combination loss was introduced to train the CNN model for cardiac segmentation in \cite{oktay2018anatomically}. Cholakkal \emph{et al.} \cite{cholakkal2019object} also proposed two special terms in their loss function to get desirable result in instance segmentation. Thus, leveraging these specific information could help improve the performance of the model. Our loss function is designed based on our proposed crossover-patch.

Our proposed method has a large difference with previous segmentation methods. Specifically, compared with the previous methods, the major distinctions of Crossover-Net are (1) our sampled patch is cross-shaped, (2) our model learns the representation from two directions simultaneously in an end-to-end manner, and (3) our loss function is able to make full use of the proposed crossover-patches.
\section{Our Method}
\label{sec:Method}
\subsection{The Architecture of Crossover-Net}
\label{subsec:architecture}
As aforementioned, our Crossover-Net includes two branches, \emph{i.e.}, vertical branch and horizontal branch, trained by crossover-patches (see Fig. \ref{framework}). Here, for CT kidney tumor segmentation, we set the size of the orthogonal patch pair to $100 \times 20$ and $20 \times 100$, respectively. Both the numbers of kernels and feature maps are experimentally determined by the rough inner cross-validation.

In particular, each branch consists of seven convolutional layers, two max pooling layers, and one linear regression layer. Also, each convolutional layer is followed by the dropout \cite{hinton2012improving} operation, the rectified linear unit (ReLU) \cite{nair2010rectified} activation and performed with stride of one and without any padding. Considering the size of our sampled patch is non-square, we set the convolutional kernels of each convolutional layer to be non-square for the consistency. The pooling operation is performed with stride of two and without any padding. The number and size of feature maps and the size of kernels in each layer are indicated in Fig. \ref{framework}. After the last convolutional operation (Fig. \ref{framework}, $conv_{7}$), the last layer in each branch becomes a patch-to-pixel mapping with 500 units being included. We concatenate the last layer of each branch directly, namely the prediction-loss layer, to combine the two branches (\emph{i.e.}, vertical branch and horizontal branch) together. The prediction-loss layer is required to be involved in the loss calculation. The purpose of this design is to capture high-level context information efficiently since features extracted by deep layers are more abstracted. Meanwhile, we also insert two crossover-loss layers (the green and blue layers in Fig. \ref{framework}) at the shallow or middle layers to capture the local information to contribute to the loss calculation. Thus, based on the prediction-loss layer and the crossover-loss layer, the last layer of the model outputs the probability of the central pixel of current crossover-patch belonging to tumor.

It is worth noting that the structure of each branch could be easily modified according to different scenarios. As for which layer can be taken as the crossover-loss layer, we introduce it in Section \ref{subsec:loss} and discuss this issue in Section \ref{sec:experiments}.

\subsection{Loss Function}
\label{subsec:loss}

Formally, our loss function integrates two terms:

\indent $\bullet$ The first one is \textbf{the prediction loss term} to measure if the prediction of Crossover-Net is correct according to the given ground truth in the training process. This term focuses on the global context, thus its corresponding prediction-loss layer is calculated on the last convolutional layer.

\indent $\bullet$ The second one is \textbf{the constraint loss term} which is calculated based on the crossover-loss layer. This term is designed according to the shape of crossover-patch. For each crossover-patch, there is an overlap region between the vertical and horizontal patches. We impose (i) the features learned by one branch from the overlap region to be similar to that of the other branch in the same layer, while (ii) the features learned from the end regions of vertical and horizontal patches are encouraged to preserve the diversity on these regions. To achieve this purpose, the spatial structural details are taken into consideration. The constraint term is designed to utilize these details. We set the corresponding crossover-loss layer on the shallow or middle layers since more local information is learned on these layers. To find which layer is desirable, we have evaluated the segmentation performance layer by layer (Section \ref{subsec:chara}). In the kidney tumor case in Fig. \ref{framework}, the middle layer in each branch, \emph{i.e.}, $conv_{3}$, is more suitable.

Thus, we define our loss function as the combination of the prediction loss $L_{pre}$ and constraint loss $L_{cs}$ as follows:
\begin{equation}
L=L_{pre}+L_{cs}.
\label{con:loss}
\end{equation}

Formally, the prediction loss term $L_{pre}$ is defined by the popularly used cross-entropy loss as follows:
\begin{equation}
L_{pre}=-\frac{1}{n}\sum_{i=1}^{n}\Bigg( y^{(i)}\text{log}\widehat{y}^{(i)}+(1-y^{(i)})\text{log}(1-\widehat{y}^{(i)})\Bigg),
\end{equation}
where $n$ is the number of training patches, $y^{(i)}$ and $\widehat{y}^{(i)}$ are the ground truth and predicted label of the central pixel in the $i$th patch, respectively.
\begin{figure}[tb]
\centering
\vspace{-0.8cm}  
\setlength{\abovecaptionskip}{-0.1cm}   
\includegraphics[width =3.2in]{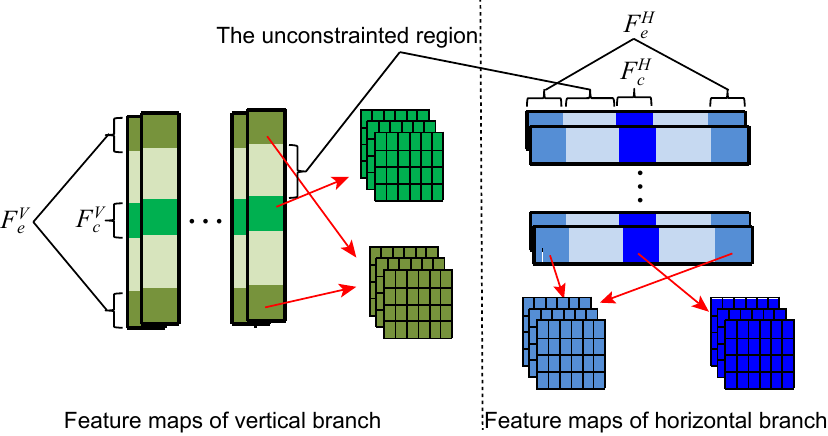}
\caption{The crossover-loss layer in vertical and horizontal branches.}\label{fig:crossloss}
\end{figure}
\begin{figure}[tb]
\centering
\vspace{-0.8cm}  
\setlength{\abovecaptionskip}{-0.1cm}   
\includegraphics[width =2.8in]{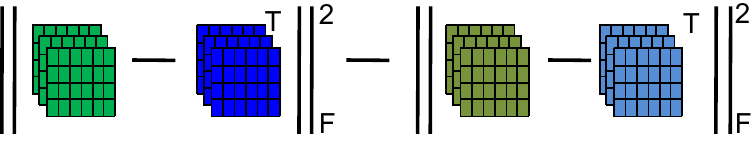}
\caption{A diagrammatic map of constraint loss term based on Fig. \ref{fig:crossloss}. The T is the transpose operator, and the $\left \| \cdot  \right \|_{F}^{2}$ is the squared Frobenius norm.}\label{fig:function}
\end{figure}

We define the constraint loss term $L_{cs}$ in Eq. (\ref{con:loss}) based on the crossover-loss layer. As shown in Fig. \ref{fig:crossloss}, we illustrate each crossover-loss layer of the two branches in detail. In the crossover-loss layer of the vertical branch, two ends of each feature map correspond to the two ends of the vertical patch in the first layer, which are denoted as $F_{e}^{V}$. Meanwhile, there is also a part which is mapped from the overlap region of crossover-patch, namely $F_{c}^{V}$. The counterparts in horizontal branch of $F_{e}^{V}$ and $F_{c}^{V}$ are denoted as $F_{e}^{H}$ and $F_{c}^{H}$, respectively. Then, we make a diagrammatic drawing for the constraint loss term in Fig. \ref{fig:function}. Specifically, for the $i$th training crossover-patch, let $\mathbf{M}_{V_{c}}^{(i)}$, $\mathbf{M}_{H_{c}}^{(i)}$, $\mathbf{M}_{V_{e}}^{(i)}$ and $\mathbf{M}_{H_{e}}^{(i)}$ denote the matrix of $F_{c}^{V}$, $F_{c}^{H}$, $F_{e}^{V}$ and $F_{e}^{H}$, respectively. We can define the $L_{cs}$ as
\begin{equation}
L_{cs} = \frac{1}{n}\sum_{i=1}^{n}\Bigg(  \left \| \mathbf{M}_{V_{c}}^{(i)}-(\mathbf{M}_{H_{c}}^{(i)})^\top \right \|_{F}^{2}  -  \left \| \mathbf{M}_{V_{e}}^{(i)}-(\mathbf{M}_{H_{e}}^{(i)})^\top \right \|_{F}^{2} \Bigg). \label{con:crossloss}
\end{equation}
\begin{figure}[tb]
\centering
\setlength{\abovecaptionskip}{-0.1cm}
\includegraphics[width =3in]{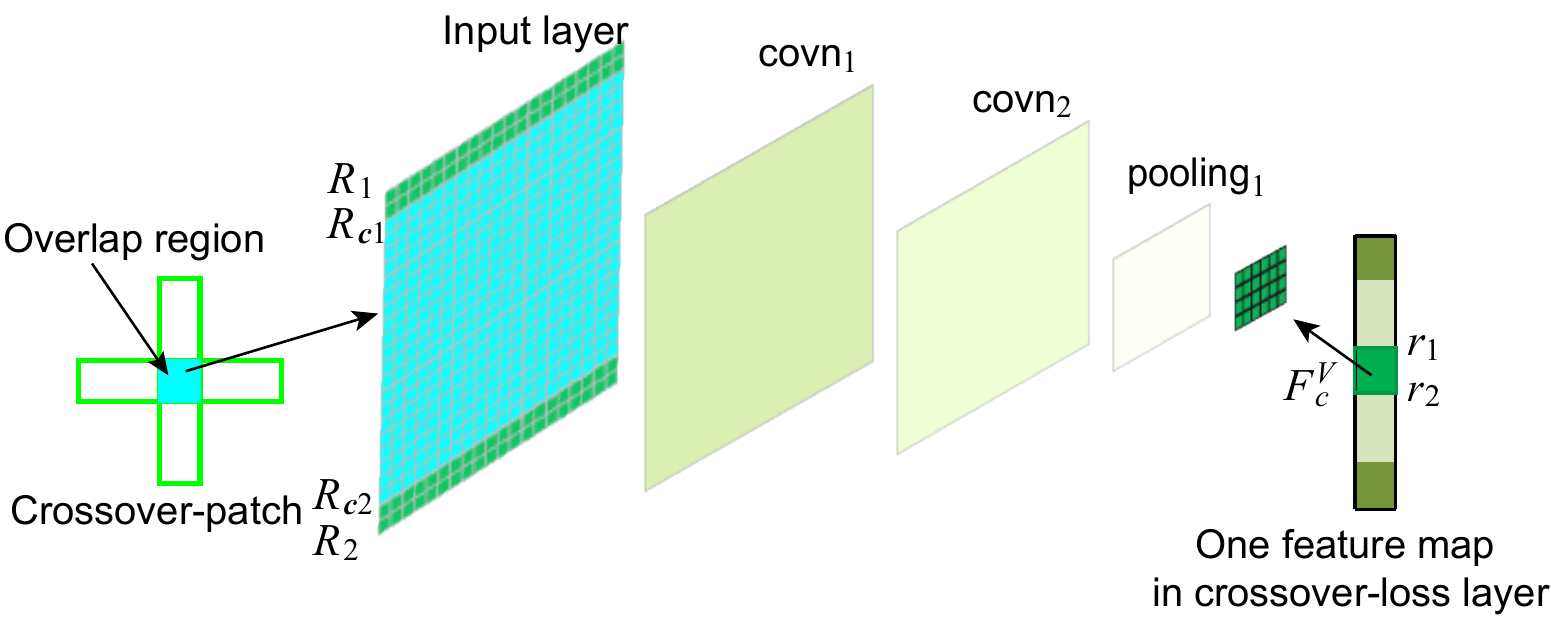}
\caption{Illustration of receptive region of $F_{c}^{V}$.}\label{fig:receptive}
\end{figure}
Now, we describe how to determine the size of the four matrices in Eq. (\ref{con:crossloss}). Taking the crossover-loss layer of vertical branch as an example here, the $F_{c}^{V}$ in Fig. \ref{fig:receptive} is the region that is influenced significantly by the overlap region of the crossover-patch (the cyan region). That is to say, in one feature map, we need to search the region whose receptive region on the input layer is closest to the overlap region. As shown in Fig. \ref{fig:receptive}, $r_{1}$ and $r_{2}$ denote the start and end row of $F_{c}^{V}$ with $R_{1}$ and $R_{2}$ being the row numbers of its receptive region, and  $R_{c1}$, $R_{c2}$ being row numbers of the overlap region. Before searching the target $F_{c}^{V}$, let we reiterate the configuration of our model: stride with $1$ and no padding in the convolutional process, and stride with $2$, filter size being $2 \times2 $ and no padding in the pooling process. Under this circumstance, assuming $r_{1}$ and $r_{2}$ are known, $R_{1}$ and $R_{2}$ can be calculated by Algorithm \ref{alg:receptive}.
\begin{algorithm}[htb]
\caption{Calculation of Receptive Region }
\label{alg:receptive}
{
\begin{algorithmic}[1] 
\REQUIRE ~~\\ 
Index of the crossover-loss layer in the branch: $l$;\\
Start and end row of $F_{c}^{V}$: $r_{1}$ and $r_{2}$;\\
Height of filters in each convolutional layer: $h$
\ENSURE ~~\\ 
Receptive region row numbers of $F_{c}^{V}$: $R_{1}$ and $R_{2}$;
\STATE $R_{1}=r_{1}, R_{2}=r_{2};$\
\FOR{each $i \in [l,1]$}
\IF{ the former layer is pooling layer}
\STATE $R_{1}=2R_{1}-1$, $R_{2}=2R_{2}$;\
\ELSE
\STATE $R_{1}=R_{1}$, $R_{2}=R_{2}+h[i]$;\
\ENDIF
\ENDFOR
\RETURN $R_{1}$, $R_{2}$; 
\end{algorithmic}
}
\end{algorithm}
The $r_{1}$ and $r_{2}$ are our targets which make the $R_{1}$ and $R_{2}$ closest to $R_{c1}$, $R_{c2}$ in Fig. \ref{fig:receptive}. Then, the size of $F_{e}^{V}$ is set the same as that of $F_{c}^{V}$. The $F_{c}^{H}$ and $F_{e}^{H}$ are obtained in the same way. Thus, in the architecture shown in Fig. \ref{framework}, the size of $\mathbf{M}_{V_{c}}^{(i)}$, $\mathbf{M}_{H_{c}}^{(i)}$, $\mathbf{M}_{V_{e}}^{(i)}$ and $\mathbf{M}_{H_{e}}^{(i)}$ are $64 \times 4 \times 6$, $64 \times 8 \times 6$ (the two $F_{e}^{V}$ in each feature map are concatenated), $64 \times 6 \times 4$ and $64 \times 6 \times 8$ (the two $F_{e}^{H}$ are concatenated), respectively. Please note that, these values could be changed according the structure of the model.
\subsection{Crossover-patch Sampling Strategy}
\label{subsec:strategy}
The performance of segmentation models in medical images largely depends on how they are aware of the boundary, however, it is usually considered to be hard to segment the tissue boundaries in practice \cite{shi2015semi}. Thus, any effort is worth doing as long as it helps to distinguish the boundary, including sampling training patches in a more effective manner. Therefore, we focus on the pixels which are distributed around the tumor boundary and set them be the centers of the crossover-patches.

\begin{figure}[tb]
\centering
\setlength{\abovecaptionskip}{-0.1cm}
\includegraphics[width =2.2in]{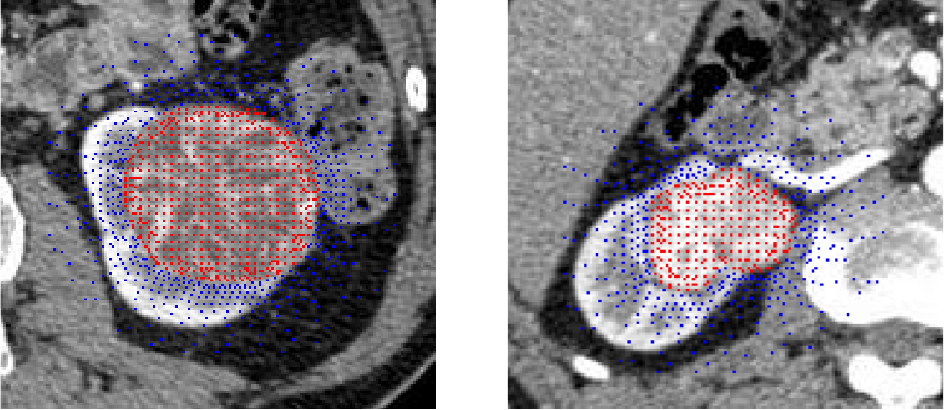}
\caption{Examples of crossover-patch sampling strategy. The red and blue pixels are centers of tumor and non-tumor patches, respectively.}\label{fig:sample}
\end{figure}

Specifically, we first select pixels densely around the boundary (both outside and inside) and uniformly inside the tumor. Then, we select pixels sparsely outside the target tissue: the further away from the boundary, the sparser the pixels are selected. In this way, the number of patches closer to the boundary is increased, and the number of those redundant patches far away from the tumor is reduced. We illustrate some typical samples in Fig. \ref{fig:sample}.

\section{Experimental Results}
\label{sec:experiments}
In this section, we extensively validate our contribution qualitatively and quantitatively. First, the datasets and the evaluation criteria are briefly introduced. Then, the characteristics of Crossover-Net are fully investigated. After that, we evaluate the performance of Crossover-Net in non-elongated tissue segmentation tasks by comparing with the state-of-the-art methods in three tasks, \ie, the kidney tumor, breast mass, and cardiac segmentation tasks. Finally, we discuss the difference between patch-based and image-based model.

\subsection{Datasets}
\textbf{Kidney Tumor.}
The abdominal CT angiographic images are acquired on a Siemens dual-source 64-slice CT scanner. Totally, 4,046 slices of 74 subjects are used in this experiment, with one or two tumors per slice. Each slice is $512 \times 512 \times L$, where $L \in [20,106]$ is the number of sampling slices along the long axis of the body. We randomly divide the data set into three parts for training, validation, and testing.  Since the data set contains many types of tumors and each type has certain characteristics, although the data sets are randomly extracted, it should be ensured that at least one case in the training set is of the same type with the test set. In addition, because kidney tumors are growing on both sides of the spine and below the front and back mid-line of the abdominal cavity, we sampled only in the second half of the abdominal cavity during the test procedure, with the left tumor being sampled on the right side of the spine and vice versa.

\textbf{Breast Mass.}
Two publicly available mammogram datasets, INbreast \cite{moreira2012inbreast, cardoso2017mass} and DDSM \cite{Heath2000digital, Heath1998current, Anmol2015} are used here.

The images in INBreast are annotated with high quality. This dataset provides 116 images with one or two masses per image and the size of each image is $3,328 \times 4,084$ or $2,560 \times 3,328$. We generate a bounding box for each mass according to their annotations. The whole dataset is divided into two mutually exclusive equal subsets for training and test. For the DDSM, there are 1,923 malignant and benign cases, and each case includes two images of each breast. ROIs are given in images containing suspicious areas. Since the ROI is not the accurate boundary of a tumor, the boundary of each tumor is annotated again as the ground truth by the experienced radiologists.

\begin{figure}[tb]
\centering
\setlength{\abovecaptionskip}{-0.1cm}
\includegraphics[width =3in]{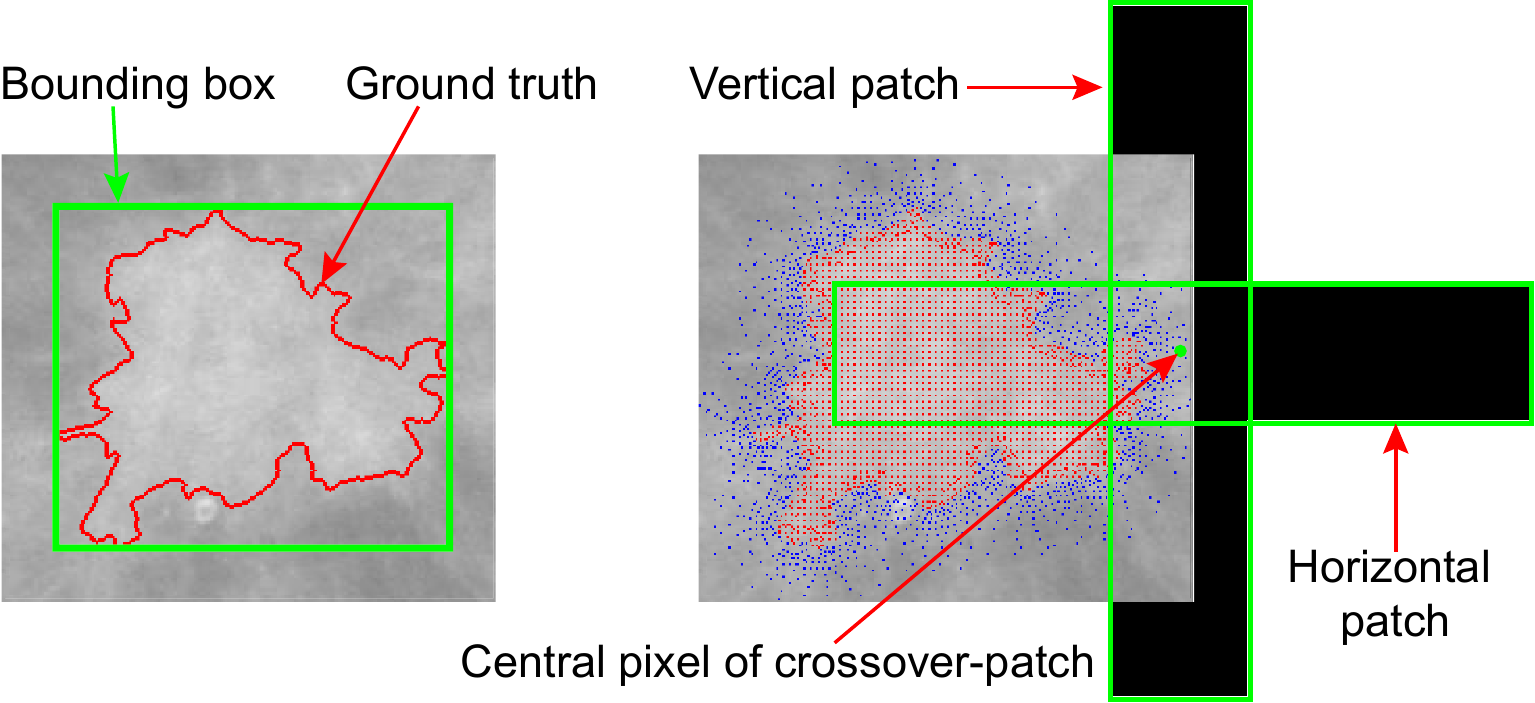}
\caption{Example of the cropped image in INBreast. The left one is the cropped image and the right one shows the crossover-patch extracted outside the image.}\label{fig:breastcropping}
\end{figure}

In most of the deep methods which segment INBreast and DDSM, each image is cropped to the bounding box \cite{dhungel2015deep, zhu2018Adversarial} or 1.2 times of the bounding box \cite{cardoso2017mass}. We crop the images in the same manner with \cite{cardoso2017mass}. If there are some crossover-patches being extracted outside the image, the outside parts are filled with black. An example is shown in Fig. \ref{fig:breastcropping}.

\textbf{Cardiac.}
We also evaluate our Crossover-Net on a public benchmark dataset of cardiac MR sequences\cite{andreopoulos2008efficient}. This dataset consists of 7,980 MR images for 33 individuals, with each image being $256 \times 256$. In the cardiac image, the boundary of the left ventricle (LV) cave (\emph{i.e.}, endocardium) and the myocardium (MYO, enclosed by endocardium and epicardium) are targets of segmentation. In each image, endocardial and epicardial contours are provided as the ground truth.

\subsection{Implementation Details}
For the scale of crossover-patch on these three datasets, we set the size of patch pair to $20 \times 100$ and  $100 \times 20$ on kidney and cardiac datasets. The structure of the model has been shown in Fig. \ref{framework}. As for the INBreast and DDSM, according to the observation about the size of masses, the crossover-patch is set to $68 \times 340$ and $340 \times 68$. For the large patches of the mammography, the depth of the model is also large. There are eleven convolutional layers, three max pooling layers, and one linear layer in each branch. The learning rate of all models is set to 0.0001. Each sub-model reaches its convergence within 16, 18, and 15 epochs on kidney, cardiac, and breast data, respectively. The training and test procedure is repeated three times in all experiments for the credibility of the segmentation. In each time, the training, validation and test sets are selected randomly in kidney and cardiac data and DDSM. Due to the limited number, there is no validation set on INBreast. We report the final average performance. All deep models are implemented on a GPU server with NVIDIA GTX 1080 Ti.

We train the Crossover-Net with the standard back propagation. The model weights are first initialized with the Xavier algorithm \cite{glorot2010understanding} which can determine the initialization scale according to the number of input and output neurons automatically \cite{wachinger2017deepnat}. In order to minimize the loss function, the parameters are updated by employing the stochastic gradient descent algorithm during the training procedure.

\subsection{Evaluation Criteria}
We employ four popular criteria to evaluate the performance of different methods, \emph{i.e.}, the Dice ratio score (DSC), the Hausdorff distance (HD), the over-segmentation ratio (OR) and the under-segmentation ratio (UR). The DSC is used to measure the overlap between final prediction and manual segmentation. The HD indicates the max min Euclidean distance between each pixel of the segmentation result and the manual result. A smaller HD indicates a higher proximity between ground truth and the segmentation result. Please refer to \cite{shi2016learning, huang20183d} for more details of these four criteria.

\subsection{Characteristics of Crossover-Net}
\label{subsec:chara}
We conduct the extensive experiment on the kidney tumor dataset to fully investigate our proposed model which include the aspects of crossover-patch, loss function and crossover-loss layer.

$\bullet$ \textbf{Advantages of Crossover-patch.}
The advantage of crossover-patches is relative to the square patches and the vertical or horizontal patches (a crossbar patch in \cite{yu2019crossbar} is essentially a separate vertical or horizontal patch since they are used separately). Similarly, one comparison object of learning from two directions is the performance of learning from single direction. Both the learning based on square patches and vertical (horizontal) patches are the manner of learning from a single direction. Therefore, we carry out the experiment with square patches, vertical and horizontal patches, and crossover-patches, respectively. In addition, simply combining the results of the vertical and horizontal patch-based networks is also a way to learn from two directions.

Specifically, the two branches of Crossover-Net are taken out to form two separate networks with the vertical and horizontal patches being their input data, respectively. Then, we set the size of square patches to $28 \times 28$, $56 \times 56$ and $100 \times 100$. The size of filters in the corresponding networks are set to $3 \times 3$, and other parameters are set the same as Crossover-Net. Here, the cross-entropy loss function is used in all networks. Also, for fair comparison, we run Crossover-Net on crossover-patches only with cross-entropy loss function.

\begin{figure}[tb]
\centering
\setlength{\abovecaptionskip}{-0.1cm}
\includegraphics[width = 3in]{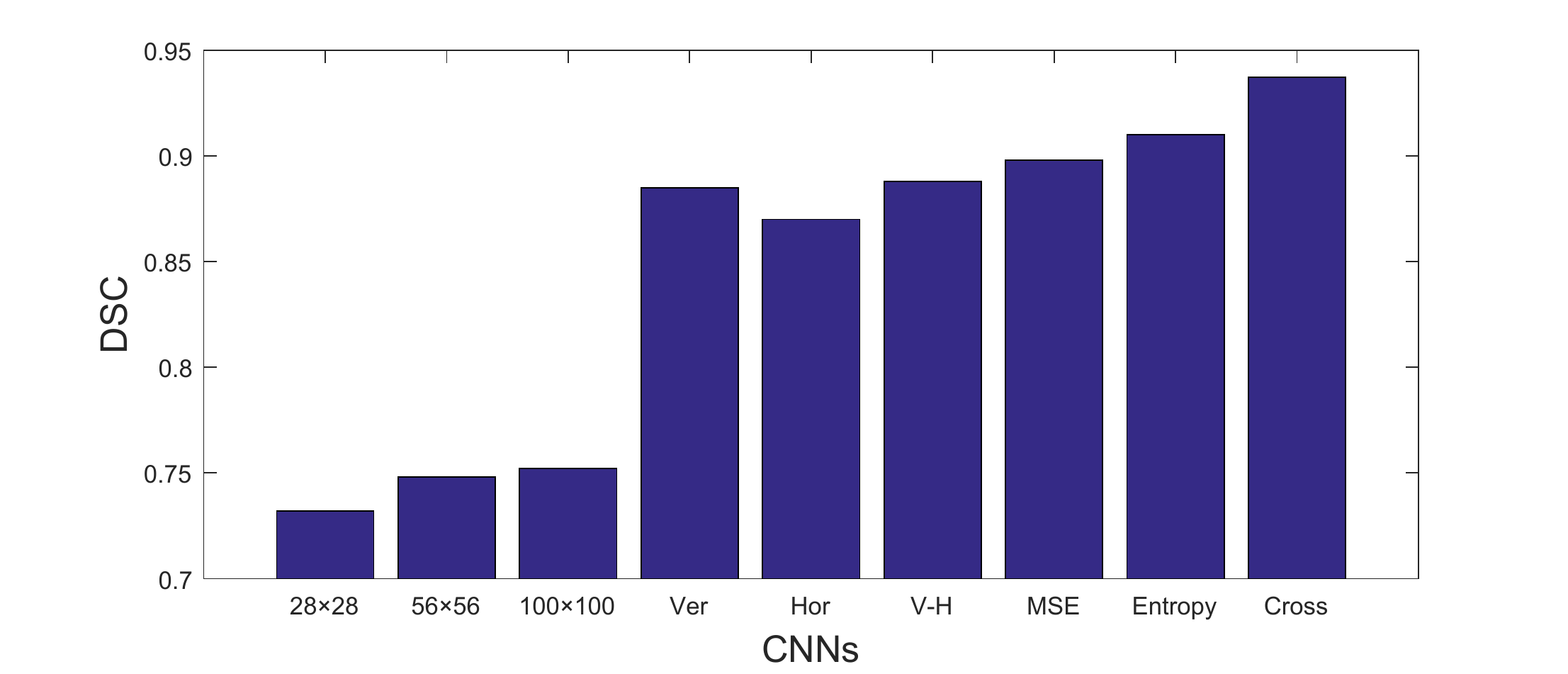}
\caption{DSC of CNNs with various patch size and loss function on kidney tumor dataset. 28, 56, 100, Ver (vertical), and Hor (horizontal) means the network based on the corresponding patch, respectively. V-H means the combination result of Ver and Hor. MSE and Entropy denotes Crossover-Net with MSE loss and cross-entropy loss, respectively. Cross means the Crossover-Net with our loss. }\label{fig:dsc}
\end{figure}

The results are shown in Fig. \ref{fig:dsc}. From the figure, we could observe that (1) the vertical network outperforms the horizontal network, indicating that the vertical features are more discriminative than the horizontal features on this dataset. (2) The DSC of the combination of the separated vertical and horizontal networks has not been improved considerably compared with the single vertical one. (3) Our Crossover-Net achieves the remarkable performance. Our patch could provide background-object-background symmetric information from long-side in orthogonal directions, leading to a complement between patch pair of the crossover-patch. Thus, it is reasonable that our model outperforms the single vertical and horizontal networks and the combination of these two models. (4) The DSCs of the three square-patch CNNs are much lower than other models. We also note that although the $100 \times 100$ patch is large enough to cover the long-side of our crossover-patch, it has not outperformed the other two small square patches significantly. This observation shows that a large patch does not mean the desirable result.

\begin{figure}[tb]
\centering
\setlength{\abovecaptionskip}{-0.1cm}
\includegraphics[width = 3in]{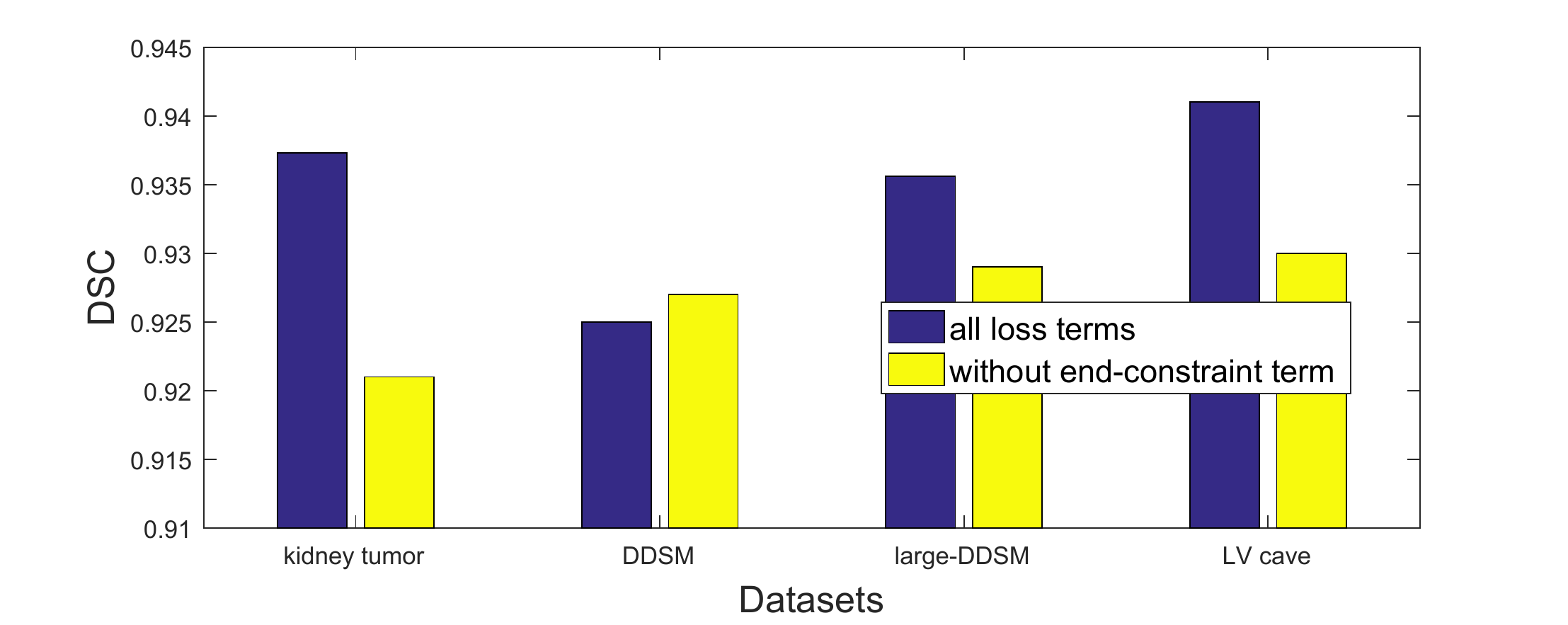}
\caption{DSC of Crossover-Net with different loss terms on three datasets. The large-DDSM means DDSM with a large crossover-patch ($50 \times 500$ and $500 \times 50$). }\label{fig:con}
\end{figure}

$\bullet$ \textbf{Advantage of Loss Function.} Next, we verify the effectiveness of the loss function. We implement our model with three types of loss functions respectively which are MSE loss, cross-entropy loss, and both cross-entropy and constraint loss. The segmentation results are shown in Fig. \ref{fig:dsc}. Obviously, the third model outperforms the former two. It also can be seen that the cross-entropy loss is slightly better for kidney tumor segmentation compared with the MSE loss. In addition, the performances of our model with different loss functions are all superior to that of the single vertical and single horizontal models. This result further indicates that learning from two directions simultaneously is better than learning from a single direction.
\begin{figure}[tb]
\centering
\setlength{\abovecaptionskip}{-0.1cm}
\includegraphics[width = 3in]{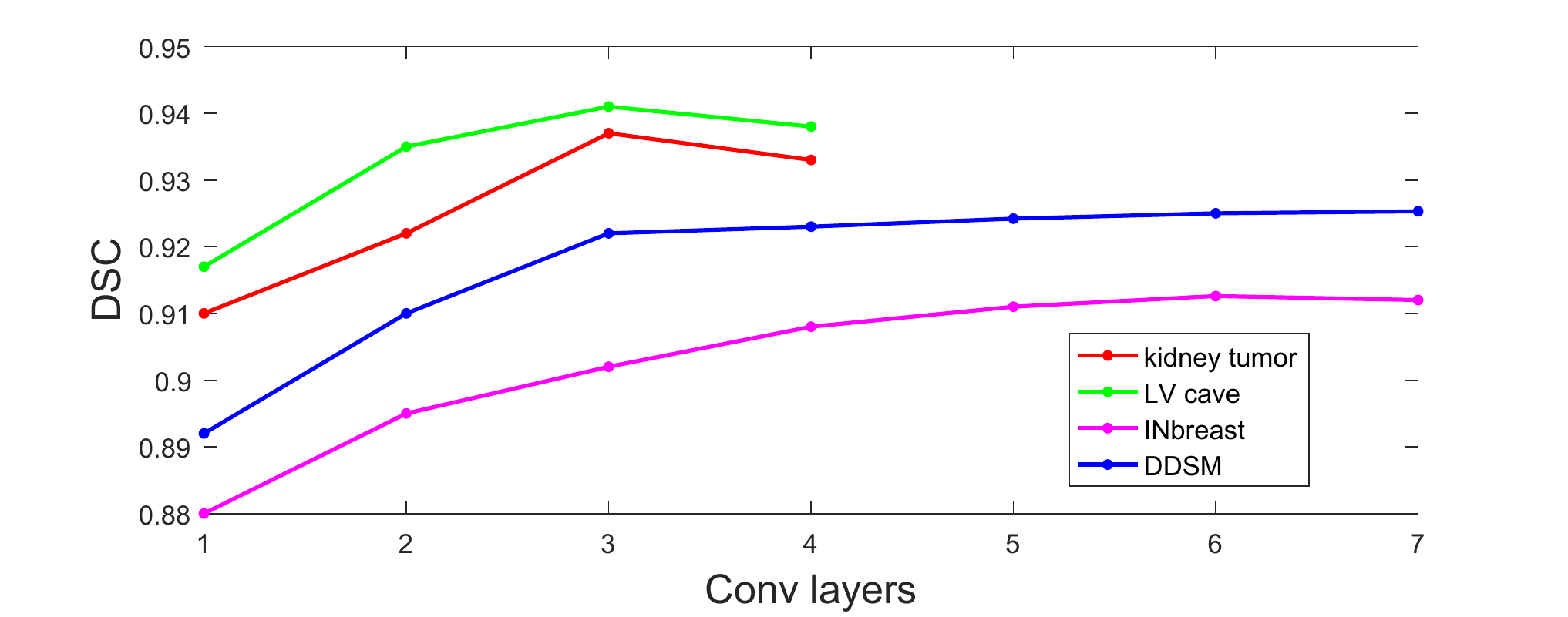}
\caption{DSC of Crossover-Net with the crossover-loss layer being set on different layers. }\label{fig:losslayer}
\end{figure}

 It is worth noting that the constraint loss also includes two terms, as shown in Eq. \ref{con:crossloss} and we name the second one as end-constraint term. The end-constraint term aims to make features learned from end regions of crossover-patch in each branch as different as possible. This is designed based on the assumption that the ends of the vertical patch are usually dissimilar with the ends of its corresponding horizontal patch. However, sometimes they might be similar. For instance, in DDSM dataset, the size of most masses is larger than the long side of our crossover-patch (340 pixels) and many benign masses are uniform in gray. Hence a large number of patches are completely contained within the mass and many vertical patches might be similar to their corresponding horizontal patches. This might lead the end-constraint term to have a negative impact on the performance of our model on this dataset. Therefore, we implement Crossover-Net with all loss terms and without the end-constraint term, respectively. As expected, the former outperforms the latter on all datasets except for DDSM (Fig. \ref{fig:con}). Nevertheless, this observation does not mean that our loss function is not suitable for the DDSM dataset. We expand the cropped region of this dataset from 1.2 times of the bounding box to 2 times. Then, we change the patch size from $68 \times 340$ and $340 \times 68$ to $50 \times 500$ and $500 \times 50$, large enough to cover most masses in long side. As is illustrated in Fig. \ref{fig:con}, the changes of DSC on DDSM with large patches is now consistent with that on the other two datasets, fully exhibiting the good fitness between our loss function and the DDSM. For this dataset, we will report the performance of our model with small patches in Section \ref{subsec:breast} for its favorable competitiveness compared with other methods specially designed for breast mass segmentation.

$\bullet$ \textbf{Advantage of Crossover-loss Layer.} The crossover-loss layer is designed to enforce the constraint loss term which concerns the low-level features. So this layer should be designed at the shallow layers. As for which layer should be chosen, we determine it by evaluating the segmentation accuracy of Crossover-Net with each convolutional layer being chosen from the lowest to the middle. Seven and eleven convolutional layers are included in our model for kidney tumor and breast mass segmentation, respectively. The cardiac dataset uses the same architecture with the kidney tumor dataset. We record the DSC of the model with the crossover-loss layer being set on different layers for these segmentation tasks (Fig. \ref{fig:losslayer}). Obviously, the accuracy is desirable when the crossover-loss layer is set on the middle layers. Both discriminative information and noise are included in the low-level features, and this might be the reason why the shallow layers cannot compete with the middle layers. In the deep layers, the degree of abstraction of features is increased. Thus, the advantage of the crossover-loss term is reduced, resulting in a decrease or no significant improvement in the segmentation accuracy.

\begin{table}
  \renewcommand\arraystretch{1.2}
  \fontsize{7.5}{8}\selectfont
  \setlength{\abovecaptionskip}{-0.1cm}
\caption{Comparison of different methods on kidney tumors}
\label{table_adnc}
\begin{center}
\begin{tabular}{c|cccc}
\toprule
&DSC&HD&OR &UR\\
\midrule
3D-FCN-PPM \cite{yang2018auto}    &0.833&12.474&0.121&0.171 \\
U-Net \cite{ronneberger2015u}     &0.839&13.020&0.108&0.169  \\
V-Net \cite{milletari2016v}       &0.885&10.440&0.065&0.152 \\
SegCaps \cite{lalonde2018capsules}&0.875&10.500&0.072&0.159\\
2PG-CNN \cite{razzak2018effi}     &0.882&11.320&0.080&0.128 \\
3D-CNN \cite{dey2018diagnostic}   &0.813&14.230&0.179&0.134 \\
Crossbar-Net                      &0.917&8.853&0.056&\textbf{0.097}\\
Crossover-Net&\textbf{0.937}&\textbf{8.627}&\textbf{0.041}&\textbf{0.097}\\
\bottomrule
\end{tabular}
\end{center}
\end{table}

\begin{figure}[tb]
\centering
\includegraphics[width = 2.6in]{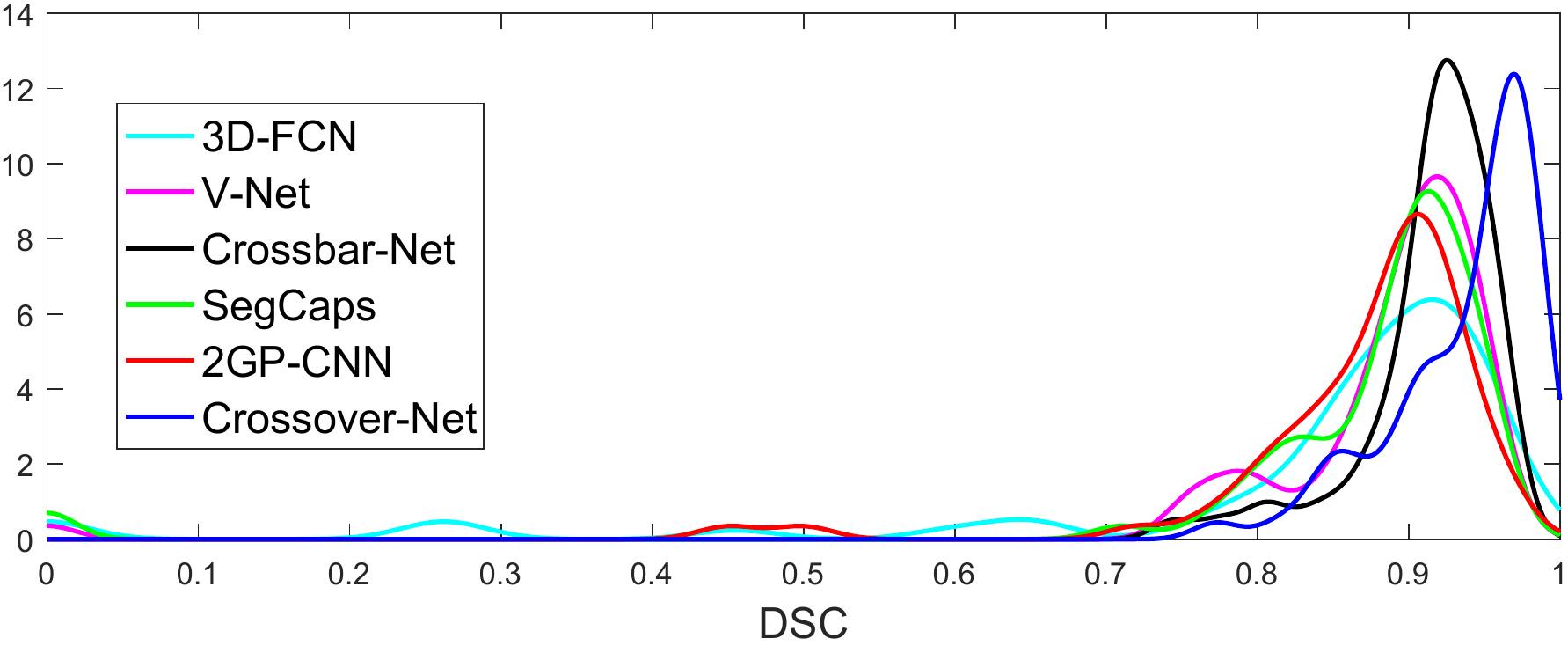}
\caption{Distribution of DSC on 600 kidney tumors.}\label{dr_bar}
\end{figure}

\begin{figure*}[hbtp]
\centering
\setlength{\abovecaptionskip}{-0.1cm}
\includegraphics[width = 6.8in]{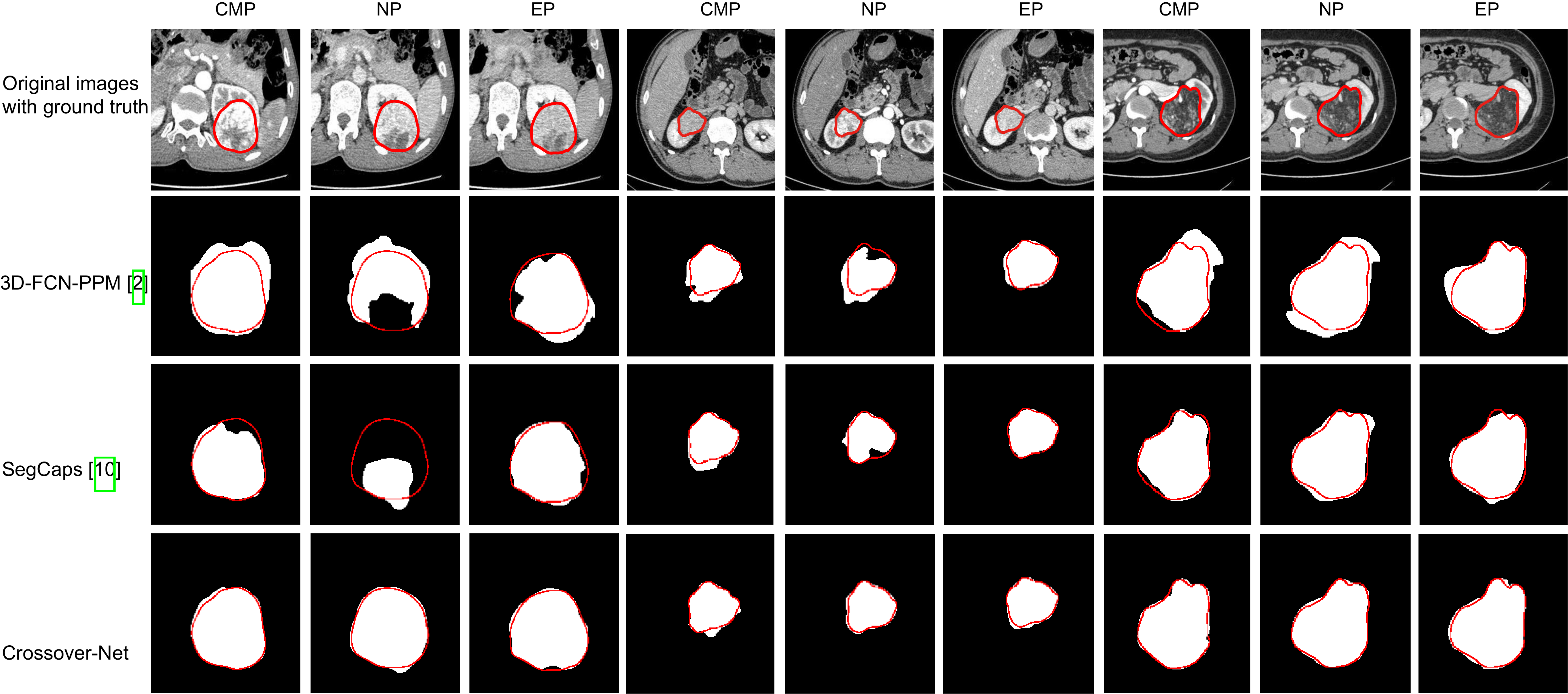}
\caption{Examples of segmentation results on the kidney tumor dataset. The red contours are the ground truth.} \label{kidney_seg_compare}
\end{figure*}

\subsection{The Results on Kidney Tumor Segmentation}
\label{sec:comparison}
First, we compare with the methods specifically designed for kidney tumor segmentation, \emph{i.e.}, \textbf{3D-FCN-PPM} \cite{yang2018auto} and our previous work \textbf{Crossbar-Net} \cite{yu2019crossbar}. In 3D-FCN-PPM, we extract ROIs manually and re-implement this 3D method as faithfully to the original manuscript as possible. We also choose several state-of-the-art methods designed for other segmentation tasks. The first is \textbf{2PG-CNN} \cite{razzak2018effi}. For this method, we refer to the code implemented in \cite{gecnn2016url}. The second is the basic \textbf{3D-CNN} \cite{dey2018diagnostic}, in which the input 3D patches are $50 \times 50 \times 5$ and $100 \times 100 \times 10$ and the size of the convolutional kernels is $3 \times 3 \times 3$. The third is \textbf{U-Net} \cite{ronneberger2015u}. The fourth is \textbf{V-Net} \cite{milletari2016v} which is a 3D segmentation model based on U-Net. The fifth is \textbf{SegCaps} \cite{lalonde2018capsules} which is a capsule network specially designed for object segmentation based on U-Net.

The quantitative results of different approaches are listed in Table \ref{table_adnc}. We have four major observations from this table. (1) The Crossover-Net provides better performance compared with other methods, with higher DSC and smaller HD, OR and UR. U-Net, 3D-FCN-PPM, and 3D-CNN exhibit similar performance. The Crossbar-Net also learns from two directions by majority voting of several sub-models, and its errors are slightly worse than those of Crossover-Net with more complex training and test procedures. (2) Although the SegCaps and 2PG-CNN are 2D methods, they perform competitively with V-Net. (3) Both 3D-CNN and 2PG-CNN are multi-scale patched-based methods, while the performance of the former, which utilizes the spatial information, is slightly inferior to that of the later. Considering that kidney tumors have a certain degree of symmetry, it is reasonable to suspect that the rotation invariance property of 2PG-CNN contributes to the good performance of the model. (4) In the aspect of under segmentation, the image-based models are slightly inferior to the patch-based methods.
\begin{table}
\fontsize{7.5}{8}\selectfont
\setlength{\abovecaptionskip}{-0.1cm}
\caption{DSC of each method in breast mass segmentation.}
\label{breast_result_metric}
\begin{center}
\renewcommand\arraystretch{1.2}
\begin{tabular}{p{1.0in}|p{0.27in}p{0.27in}}
\toprule
Method&INbreast &DDSM \\
\midrule
Dhungel \emph{et al.} \cite{dhungel2015deep}&0.8800 &0.8700 \\
Crossbar-Net \cite{yu2019crossbar} &0.9013&0.9122\\
Cross-sensor\cite{cardoso2017mass} &0.9000&0.9000\\
AM-FCN \cite{zhu2018Adversarial} & 0.9097 & 0.9130\\
Zhang \emph{et al.} \cite{zhang2018photoacoustic} & - & 0.9118\\
Crossover-Net & \textbf{0.9126} & \textbf{0.9250}\\
\bottomrule
\end{tabular}
\end{center}
\vspace{-0.2cm}
\end{table}

\begin{figure}[tb]
\centering
\setlength{\abovecaptionskip}{-0.1cm}
\includegraphics[width =3in]{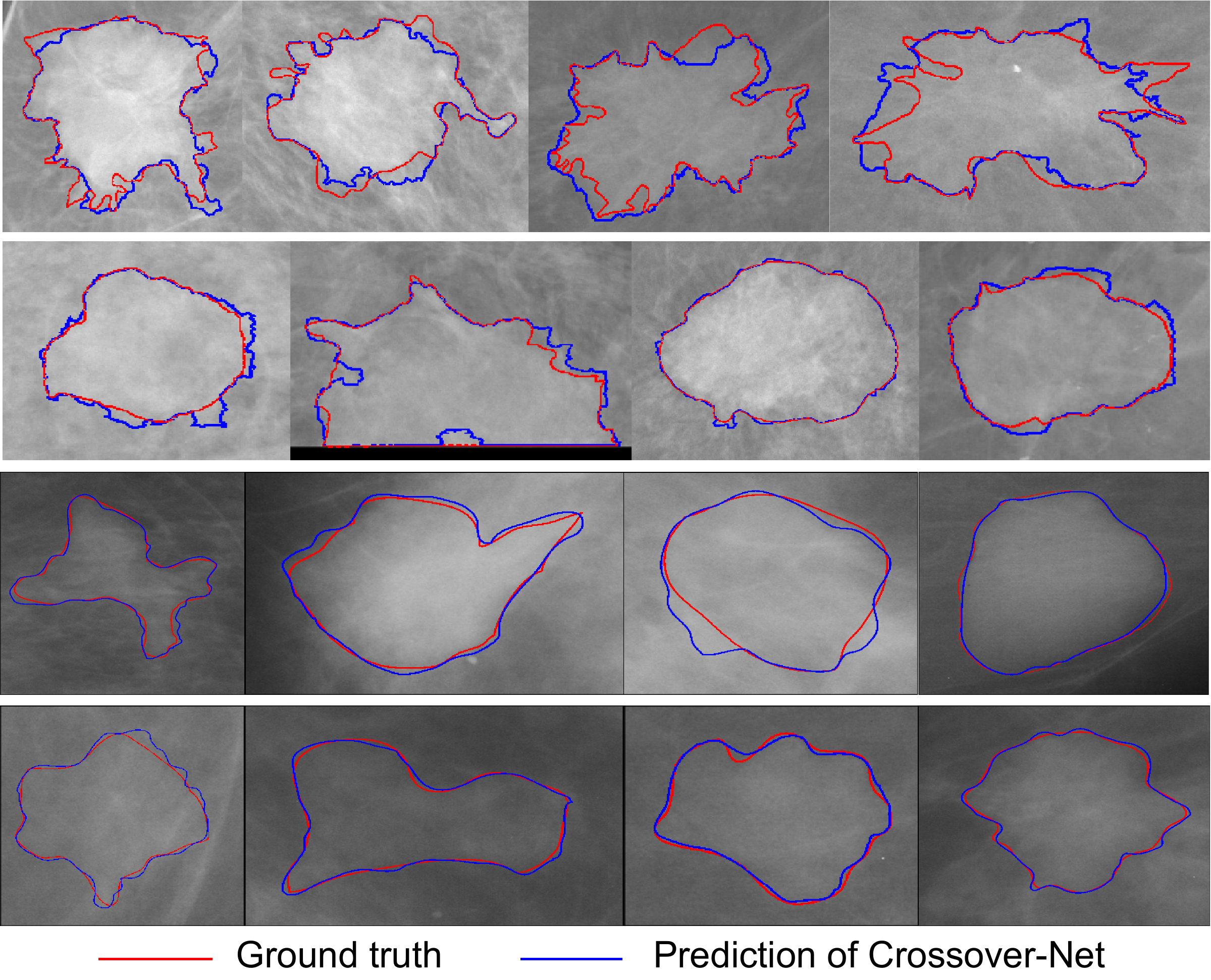}
\caption{Typical segmentation results of Crossover-Net on INbreast and DDSM. The first two rows are images in INbreast and the remaining images belong to DDSM.}\label{fig:inbreastDDSM}
\end{figure}

\begin{table}[tp]
  \centering
  \renewcommand\arraystretch{1.2}
  \fontsize{7.5}{8}\selectfont
  \setlength{\abovecaptionskip}{-0.1cm}
  \caption{Performance comparison of state-of-the-art methods on cardiac segmentation}
  \label{tab:performance_comparison}
    \begin{tabular}{c|cccc}
    \toprule
    \multirow{2}{*}{Method}&
    \multicolumn{2}{c}{LV}&\multicolumn{2}{c}{MYO}\\
    \cmidrule(lr){2-3} \cmidrule(lr){4-5}
    &DSC & HD & DSC & HD\\
    \midrule
    Li \emph{et al.} \cite{li2019fully}          &0.942&6.641&0.892&8.786\\
    Duan \emph{et al.} \cite{Duan2019Auto}       &0.943&4.09&0.854&4.37\\
    Du \emph{et al.} \cite{du2019card}           &\textbf{0.960}&-&0.890&-\\
    3D-CNN\cite{patravali20172d-3d} &0.925&14.650&0.855&38.120\\
    GridNet\cite{zotti2017gridnet}  &0.955&5.850&0.885&8.010\\
    Crossover-Net&0.941&\textbf{3.520} &\textbf{0.916}&\textbf{4.100}\\
    \bottomrule
    \end{tabular}
\end{table}

To fully observe the detailed segmentation of each method, we randomly sample 600 images from one test set. The kernel density estimation of DSC of all compared methods on these images is depicted in Fig. \ref{dr_bar}. In this figure, all the 3D-FCN-PPM, V-Net, and SegCaps have some cases of low or even 0 DSC, indicating that some tumors are under-segmented with some degree or missed by these methods. The tumors less than 15 pixels in diameter account for the vast majority of all tumors which are missed or under-segmented. Also, there are some tumors being under-segmented by 2PG-CNN (DSC is among 0.4 to 0.6), while 0 DSC case does not exist. This indicates that 2PG-CNN can detect tumors correctly whereas the features learned by this model are not discriminative enough. We visualize some typical segmentation examples of Crossover-Net, 3D-FCN-PPM, and SegCaps in Fig. \ref{kidney_seg_compare}. For each tumor, the three phases (\emph{i.e.}, CMP, EP and NP) are listed. Obviously, even the tumors vary greatly between different phases and different subjects, the predictions of Crossover-Net are still close to the ground truth.

\begin{figure*}[!tp]
\centering
\setlength{\abovecaptionskip}{-0.1cm}
\includegraphics[width = 6.2in]{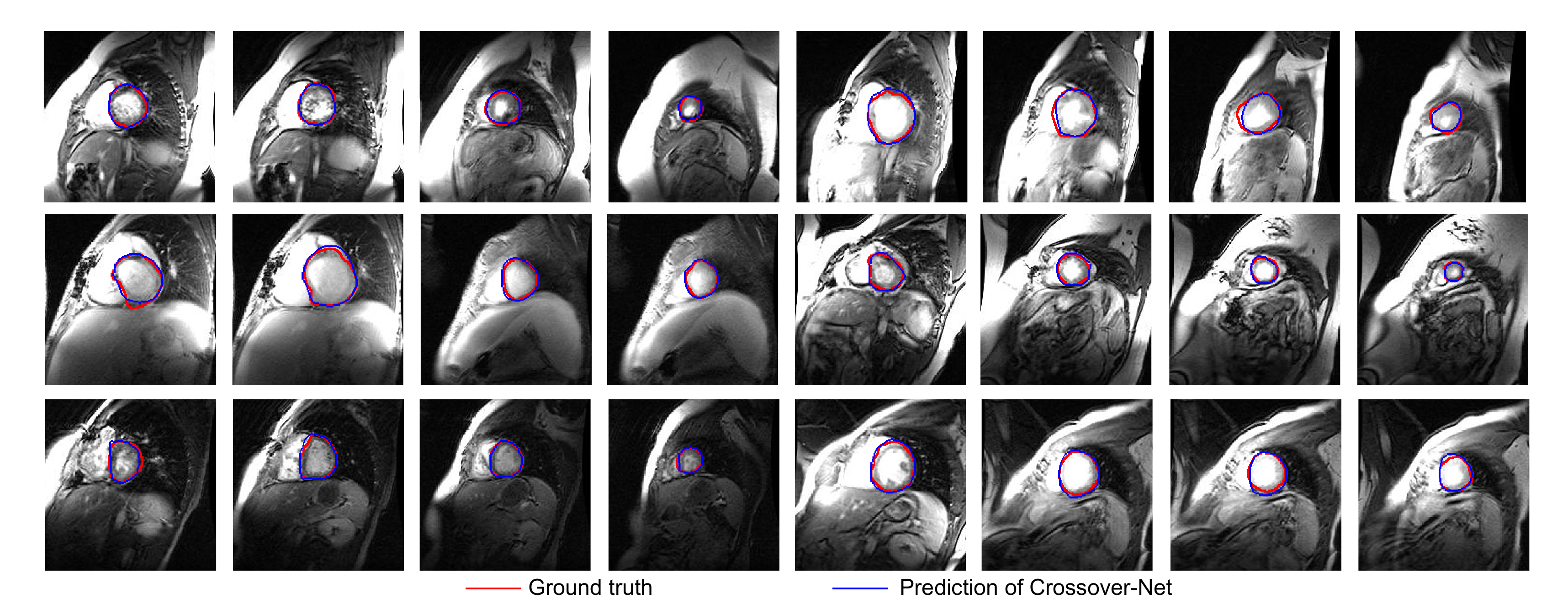}
\caption{Typical results on epicardial segmentation. }\label{heart_seg_result}
\end{figure*}

\begin{table}[tp]
  \centering
  \renewcommand\arraystretch{1.2}
  \fontsize{7.5}{8}\selectfont
  \setlength{\abovecaptionskip}{-0.1cm}
  \caption{Comparisons between image-based methods and Crossover-Net on kidney tumor segmentation.}
  \label{tab:discussion_comparison}
    \begin{tabular}{c|ccc}
    \toprule
    \multirow{2}{*}{ }&
    \multicolumn{2}{c}{2D image-based methods}&\multirow{2}{*}{Crossover-Net}\\
    \cmidrule(lr){2-3}
    &U-Net \cite{ronneberger2015u}&SegCaps \cite{lalonde2018capsules}\\
    \midrule
    Training time&$\sim$45 minutes&$\sim$27 hours&$\sim$30 minutes\\
    Storage size&139,191KB&5,628KB&1,427KB\\
    \bottomrule
    \end{tabular}
\end{table}

\subsection{The Results on Breast Mass Segmentation}
\label{subsec:breast}
The Crossbar-Net \cite{yu2019crossbar} and other four state-of-the-art methods designed especially for breast mass segmentation are employed for comparison. We take the DSC values of \cite{dhungel2015deep, cardoso2017mass, zhu2018Adversarial, zhang2018photoacoustic} from their original paper. As is shown in Table \ref{breast_result_metric}, Crossover-Net is competitive with other methods on both datasets. We also illustrate some visual examples in Fig. \ref{fig:inbreastDDSM}. The results indicate that the prediction of our method are closest to the ground truth. Although there are some masses with burring boundary (the first row in Fig. \ref{fig:inbreastDDSM}), our Crossover-Net still performs well. Besides, just as mentioned in Section \ref{subsec:chara}, many masses in DDSM are over 400 pixels in diameter, larger than the long side of the crossover-patches (340 pixels). Thus, the symmetry property of masses could not be fully captured by patches, while the segmentation results are still desirable. So, the performance of Crossover-Net is quite robust.

\subsection{The Results on Cardiac Segmentation}
\label{subsec:cardiac}
The quantified results of Crossover-Net and five state-of-the-art methods \cite{patravali20172d-3d, zotti2017gridnet, li2019fully, Duan2019Auto, du2019card} on LV cave and MYO are listed in Table \ref{tab:performance_comparison}. All the five methods are image-based methods and applied on different datasets. We report the best records of these methods according to their original literatures.

We would analyze the potential phenomena rather than simply compare the evaluation criteria. Firstly, although several methods have been proposed over the past two years, the segmentation results have not improved significantly. Secondly, the image-based method occupies a major position in cardiac segmentation, and mostly uses a coarse-to-fine model. Generally, coarse-to-fine models are usually used in the small target segmentation tasks for the small area fraction and a variable shape, such as pancreas segmentation. Unlike the pancreas, the fraction of the heart in the $256 \times 256$ image is not very small, and the shape of the heart among each subject is relatively regular. With these advantages, the 2-stage method is still used and the performance is not greatly improved. This phenomenon can only indicate that the boundary of the target is difficult to segment. Then, between the endocardium and the myocardium, which is the hard one?  As is illustrated in Table \ref{tab:performance_comparison}, the DSC of the LV in each method is generally higher than that of the MYO. Considering that the segmentation accuracy of MYO is together determined by that of endocardium and epicardium while the DSC on LV cave is decided by endocardial segmentation, we can get that it is the boundary of the epicardium that is hard to segment. Naturally, the local details will be involved when the boundary information is mentioned. Once again, the image-based methods lose some local information, especially the boundary details, leading to a large HD. Since the datasets of the various methods are not the same, it is not very meaningful to simply compare the values of DSC and HD. However,  although the performance of Crossover-Net is not the best one, our model has a closest DSC between LV and MYO, which indicates Crossover-Net has a relatively good performance on the epicardium segmentation. Therefore, as an end-to-end single-step model, Crossover-Net has a superior learning ability on boundary detail information than the imaged-based 2-stage models.

Also, we show some typical results on epicardial segmentation of eight subjects in Fig. \ref{heart_seg_result}, with four images for each subject. It can be seen that our segmentation results could fit the ground truth accurately.

\subsection{Discussion}

Since the FCN model was proposed, the image-based models have become the mainstream methods in segmentation fields, and the patch-based models tend to be gradually replaced by image-based methods. Although the patch-based models still take some role in the field of medical image segmentation, less and less work has been published in the past two years. Different from natural image segmentation, we usually have higher requirements for the accuracy of the tissue boundary segmentation. So, pixel-level segmentation/delineation is greatly required in medical image segmentation. The image-based model inevitably brings a lot of interference and noise while introducing rich contextual information. We have verified through extensive experiments that such models are easy to under-segment or ignore the small tumor and lose the boundary detail information. On the contrary, the patch-based models can help us to capture the pixel-centered information, hence we could expect them to meet our high requirements on tissue boundaries. Therefore, the patch-based methods should not be abounded in the medical image segmentation field.

 Usually, the computational cost is the main aspect of the patch-based models being questioned. We have taken corresponding measures to reduce the cost in our model. For example, in sampling strategy (Section \ref{subsec:strategy}), only patches with discriminative information are selected for training, with redundancy being reduced greatly. Additionally, for tissues like kidney tumor or left ventricle, there is always a sequence of images included in each case. In each sequence, the difference between adjacent images is generally small, hence we could sample crossover-patches every other images. Thus, the training data could be reduced by half, further reducing redundancy. In contrast to our Crossover-Net, the image-based methods not only require all images of each sequence to participate in training, but also need to augment the training data in the way of translating, rotating, reflecting, \emph{etc}, otherwise the model will be over-fitting. But we cannot guarantee that these augmentation operations will definitely make the model better. We have tried to segment the INBreast dataset using U-Net. Since only 116 images in this dataset, the training data must be augmented. However, when we added the rotation data, the DSC decreased by 0.3. In addition, in test procedure, we set sampling ranges based on the characteristics of the human body, meanwhile, we sample one pixel every three to predict its label. The labels of un-sampled pixels are determined by its neighbors. Therefore, the computing cost of our method is competitive to the image-based methods.

Taking the kidney tumor segmentation as an example, we compare Crossover-Net with U-Net and SegCaps on the training time and storage size (since the sampling region is set in the test procedure, the test time will not be compared). We implement Crossover-Net and U-Net with Python 3.6 \cite{van2009python}. For U-Net, the Dice loss is adopted and the best test results are obtained after 28 epochs. We obtain the code of SegCaps from \cite{segcaps2018code} and modify the code relevant to reading and converting images. This model is trained on four GPUs and converges after 16 epochs with one epoch taking about 100 minutes. The training time of Crossover-Net is significantly desirable (Table \ref{tab:discussion_comparison}). Also, the model storage size of our method is smaller than that of the other two methods.

Generally, for the patch-based models, learning necessary information on patches is one of the core-strengths of the convolutional network, no matter the shape of the patch is square or rectangle. Then, is it still desirable to employ the crossover-patch? We have done extensive comparison among small and large square patches, rectangular patches, and our crossover-patches on kidney tumor segmentation. The performances of square patches are obviously inferior to that of our patches. For tissues with complicated texture, the small square patch cannot capture discriminative information enough, while the large one may include more noise. We also compare with the model based on multi-scale square patches, and Table \ref{table_adnc} illustrates the remarkable performance of our model. The vertical or horizontal patches are also uncompetitive with crossover-patches obviously (see Fig. \ref{fig:dsc}).
\section{Conclusion}
\label{sec:conclusion}
In this paper, we proposed a Crossover-Net model for non-elongated tissue segmentation tasks. Three factors contribute to the superior performance of our model to learn tissue-sensitive features automatically: (1) both vertical and horizontal contextual information is included in the crossover-patch as a whole for better performance, (2) the proposed Crossover-Net tries to learn features based on the cross information in an end-to-end manner, and (3) the constraint loss term emphasizes the connection between the vertical and horizontal patches in a same crossover-patch to utilize information more effectively and efficiently. Besides, we sampled patches densely around boundaries and sparsely far away from the targets, which could effectively guarantee the distribution of patches and avoiding redundancy. The new model was applied to the segmentation task on three different modal datasets, and all achieved good segmentation results. In the experiments, we not only fully verified the good and robust performance of Crossover-Net, but also highlighted the ability of the patch-based method to capture boundary detail information by using pixel-centric local information, which is what the image-based model lacks. Finally, we also conducted in-depth discussions on the training time and storage size of the two types of models, verifying the competitiveness of our model in these two aspects compared to the image-based model.

 The problem with the new model is that when the segmentation target is too large, it is necessary to sample larger patches and design a deeper network. This will lead to an increase in training time, so the next work is to design a more efficient model. How to use cross information more effectively is the key point of future work. Furthermore, extending our model to elongated tissue segmentation tasks, such as vessel segmentation, is also an appealing topic.



%
%
%
%

\ifCLASSOPTIONcaptionsoff
  \newpage
\fi



%

\bibliographystyle{IEEEtran}
\bibliography{egpaper_final}

\begin{thebibliography}{10}
\providecommand{\url}[1]{#1}
\csname url@samestyle\endcsname
\providecommand{\newblock}{\relax}
\providecommand{\bibinfo}[2]{#2}
\providecommand{\BIBentrySTDinterwordspacing}{\spaceskip=0pt\relax}
\providecommand{\BIBentryALTinterwordstretchfactor}{4}
\providecommand{\BIBentryALTinterwordspacing}{\spaceskip=\fontdimen2\font plus
\BIBentryALTinterwordstretchfactor\fontdimen3\font minus
  \fontdimen4\font\relax}
\providecommand{\BIBforeignlanguage}[2]{{%
\expandafter\ifx\csname l@#1\endcsname\relax
\typeout{** WARNING: IEEEtran.bst: No hyphenation pattern has been}%
\typeout{** loaded for the language `#1'. Using the pattern for}%
\typeout{** the default language instead.}%
\else
\language=\csname l@#1\endcsname
\fi
#2}}
\providecommand{\BIBdecl}{\relax}
\BIBdecl

\bibitem{yang2018auto}
G.~Yang, G.~Li, T.~Pan, Y.~Kong, J.~Wu, H.~Shu \emph{et~al.}, ``Automatic
  segmentation of kidney and renal tumor in {CT} images based on {3D} fully
  convolutional neural network with pyramid pooling module,'' in
  \emph{International Conference on Pattern Recognition}, 2018, pp. 3790--3795.

\bibitem{zhu2018Adversarial}
W.~Zhu, X.~Xiang, T.~D~Tran, G.~Hager, and X.~Xie, ``Adversarial deep
  structured nets for mass segmentation from mammograms,'' \emph{IEEE
  International Symposium on Biomedical Imaging}, pp. 847--850, 2018.

\bibitem{zhang2018photoacoustic}
J.~Zhang, B.~Chen, M.~Zhou, H.~Lan, and F.~Gao, ``Photoacoustic image
  classification and segmentation of breast cancer: a feasibility study,''
  \emph{IEEE Access}, vol.~7, pp. 5457--5466, 2019.

\bibitem{jiang2019multiple}
J.~Jiang, Y.~Hu, C.~Liu, D.~Halpenny, M.~D. Hellmann, J.~O. Deasy
  \emph{et~al.}, ``Multiple resolution residually connected feature streams for
  automatic lung tumor segmentation from {CT} images,'' \emph{IEEE Transactions
  on Medical Imaging}, vol.~38, no.~1, pp. 134--144, 2019.

\bibitem{khened2018densely}
M.~Khened, V.~Alex, and G.~Krishnamurthi, ``Densely connected fully
  convolutional network for short-axis cardiac cine {MR} image segmentation and
  heart diagnosis using random forest,'' \emph{International Workshop on
  Statistical Atlases and Computational Models of the Heart}, pp. 140--151,
  2018.

\bibitem{patravali20172d-3d}
J.~Patravali, S.~Jain, and S.~Chilamkurthy, ``{2D-3D} fully convolutional
  neural networks for cardiac {MR} segmentation,'' \emph{International Workshop
  on Statistical Atlases and Computational Models of the Heart}, pp. 130--139,
  2018.

\bibitem{ronneberger2015u}
O.~Ronneberger, P.~Fischer, and T.~Brox, ``U-{Net}: Convolutional networks for
  biomedical image segmentation,'' in \emph{Medical Image Computing and
  Computer Assisted Intervention}, 2015, pp. 234--241.

\bibitem{lalonde2018capsules}
R.~LaLonde and U.~Bagci, ``Capsules for object segmentation,'' \emph{arXiv
  preprint arXiv:1804.04241}, 2018.

\bibitem{yu2019crossbar}
Q.~Yu, Y.~Shi, J.~Sun, Y.~Gao, J.~Zhu, and Y.~Dai, ``Crossbar-{Net}: A novel
  convolutional neural network for kidney tumor segmentation in {CT} images,''
  \emph{IEEE Transactions on Image Processing}, vol.~28, no.~8, pp. 4060--4074,
  2019.

\bibitem{long2015fully}
J.~Long, E.~Shelhamer, and T.~Darrell, ``Fully convolutional networks for
  semantic segmentation,'' pp. 3431--3440, 2015.

\bibitem{li2018h-denseunet:}
X.~Li, H.~Chen, X.~Qi, Q.~Dou, C.~Fu, and P.~Heng, ``H-{D}ense{UN}et: Hybrid
  densely connected {UN}et for liver and tumor segmentation from {CT}
  volumes,'' \emph{IEEE Transactions on Medical Imaging}, vol.~37, no.~12, pp.
  2663--2674, 2018.

\bibitem{ciresan2012deep}
D.~C. Cire\c{s}an, A.~Giusti, L.~M. Gambardella, and J.~Schmidhuber, ``Deep
  neural networks segment neuronal membranes in electron microscopy images,''
  \emph{International Conference on Neural Information Processing Systems},
  vol.~2, pp. 2843--2851, 2012.

\bibitem{wang2017central}
S.~Wang, M.~Zhou, Z.~Liu, Z.~Liu, D.~Gu, Y.~Zang \emph{et~al.}, ``Central
  focused convolutional neural networks: Developing a data-driven model for
  lung nodule segmentation,'' \emph{Medical Image Analysis}, vol.~40, pp.
  172--183, 2017.

\bibitem{shi2017does}
Y.~Shi, W.~Yang, Y.~Gao, and D.~Shen, ``Does manual delineation only provide
  the side information in {CT} prostate segmentation?'' pp. 692--700, 2017.

\bibitem{havaei2017brain}
M.~Havaei, A.~Davy, D.~Warde-Farley, A.~Biard, A.~Courville, Y.~Bengio
  \emph{et~al.}, ``Brain tumor segmentation with deep neural networks,''
  \emph{Medical Image Analysis}, vol.~35, pp. 18--31, 2017.

\bibitem{razzak2018effi}
I.~Razzak, M.~Imran, and G.~Xu, ``Efficient brain tumor segmentation with
  multiscale two-pathway-group conventional neural networks,'' \emph{IEEE
  Journal of Biomedical and Health Informatics}, vol.~23, no.~5, pp.
  1911--1919, 2019.

\bibitem{moeskops2016automatic}
P.~Moeskops, M.~A. Viergever, A.~M. Mendrik, L.~S. de~Vries, M.~J. Benders, and
  I.~I{\v{s}}gum, ``Automatic segmentation of {MR} brain images with a
  convolutional neural network,'' \emph{IEEE Transactions on Medical Imaging},
  vol.~35, no.~5, pp. 1252--1261, 2016.

\bibitem{kamnitsas2017efficient}
K.~Kamnitsas, C.~Ledig, V.~F. Newcombe, J.~P. Simpson, A.~D. Kane, D.~K. Menon
  \emph{et~al.}, ``Efficient multi-scale {3D CNN} with fully connected {CRF}
  for accurate brain lesion segmentation,'' \emph{Medical Image Analysis},
  vol.~36, pp. 61--78, 2017.

\bibitem{zhou2018high}
Z.~Zhou, Y.~Wang, J.~Yu, Y.~Guo, W.~Guo, and Y.~Qi, ``High spatial–temporal
  resolution reconstruction of plane-wave ultrasound images with a multichannel
  multiscale convolutional neural network,'' \emph{IEEE Transactions on
  Ultrasonics Ferroelectrics and Frequency Control}, vol.~65, no.~11, pp.
  1983--1996, 2018.

\bibitem{ng2018deep}
N.~Bien, P.~Rajpurkar, R.~L. Ball, J.~Irvin, A.~Park, E.~Jones \emph{et~al.},
  ``Deep-learning-assisted diagnosis for knee magnetic resonance imaging:
  Development and retrospective validation of {MRNet},'' \emph{PLOS Medicine},
  vol.~15, no.~11, pp. 1--19, 2018.

\bibitem{lin2018multi-scale}
D.~Lin, Y.~Ji, D.~Lischinski, D.~Cohenor, and H.~Huang, ``Multi-scale context
  intertwining for semantic segmentation.'' \emph{European Conference on
  Computer Vision}, pp. 622--638, 2018.

\bibitem{huang20183d}
Y.~Huang, Q.~Dou, Z.~Wang, L.~Liu, Y.~Jin, C.~Li \emph{et~al.}, ``3{D}
  {ROI}-aware {U}-{N}et for accurate and efficient colorectal tumor
  segmentation,'' \emph{arXiv preprint arXiv:1806.10342v5}, 2019.

\bibitem{lin2019zigzagnet:}
D.~Lin, D.~Shen, S.~Shen, Y.~Ji, D.~Lischinski, D.~Cohenor \emph{et~al.},
  ``Zig{ZagNet}: Fusing top-down and bottom-up context for object
  segmentation,'' \emph{IEEE Conference on Computer Vision and Pattern
  Recognition}, pp. 7490--7499, 2019.

\bibitem{orsic2019in}
M.~Orsic, I.~Kreso, P.~Bevandic, and S.~Segvic, ``In defense of pre-trained
  {ImageNet} architectures for real-time semantic segmentation of road-driving
  images,'' \emph{IEEE Conference on Computer Vision and Pattern Recognition},
  pp. 12\,606--12\,616, 2019.

\bibitem{li2019dfanet:}
H.~Li, P.~Xiong, H.~Fan, and J.~Sun, ``{DFANet}: Deep feature aggregation for
  real-time semantic segmentation,'' \emph{IEEE Conference on Computer Vision
  and Pattern Recognition}, pp. 9522--9531, 2019.

\bibitem{xiao2018unified}
T.~Xiao, Y.~Liu, B.~Zhou, Y.~Jiang, and J.~Sun, ``Unified perceptual parsing
  for scene understanding,'' \emph{European Conference on Computer Vision}, pp.
  432--448, 2018.

\bibitem{tian2019decoders}
Z.~Tian, T.~He, C.~Shen, and Y.~Yan, ``Decoders matter for semantic
  segmentation: Data-dependent decoding enables flexible feature aggregation,''
  \emph{IEEE Conference on Computer Vision and Pattern Recognition}, pp.
  3126--3135, 2019.

\bibitem{milletari2016v}
F.~Milletari, N.~Navab, and S.~Ahmadi, ``{V-N}et: Fully convolutional neural
  networks for volumetric medical image segmentation,'' in \emph{International
  Conference on 3D Vision}, 2016, pp. 565--571.

\bibitem{oktay2018anatomically}
O.~Oktay, E.~Ferrante, K.~Kamnitsas, M.~P. Heinrich, W.~Bai, J.~Caballero
  \emph{et~al.}, ``Anatomically constrained neural networks ({ACNNs}):
  Application to cardiac image enhancement and segmentation,'' \emph{IEEE
  Transactions on Medical Imaging}, vol.~37, no.~2, pp. 384--395, 2018.

\bibitem{cholakkal2019object}
H.~Cholakkal, G.~Sun, F.~S. Khan, and L.~Shao, ``Object counting and instance
  segmentation with image-level supervision.'' \emph{IEEE Conference on
  Computer Vision and Pattern Recognition}, pp. 12\,397--12\,405, 2019.

\bibitem{hinton2012improving}
G.~E. Hinton, N.~Srivastava, A.~Krizhevsky, I.~Sutskever, and R.~R.
  Salakhutdinov, ``Improving neural networks by preventing co-adaptation of
  feature detectors,'' \emph{arXiv preprint arXiv:1207.0580}, 2012.

\bibitem{nair2010rectified}
V.~Nair and G.~E. Hinton, ``Rectified linear units improve restricted
  {B}oltzmann machines,'' in \emph{International Conference on Machine
  Learning}, 2010, pp. 807--814.

\bibitem{shi2015semi}
Y.~Shi, Y.~Gao, S.~Liao, D.~Zhang, Y.~Gao, and D.~Shen, ``Semi-automatic
  segmentation of prostate in {CT} images via coupled feature representation
  and spatial-constrained transductive {L}asso,'' \emph{IEEE Transactions on
  Pattern Analysis and Machine Intelligence}, vol.~37, no.~11, pp. 2286--2303,
  2015.

\bibitem{moreira2012inbreast}
I.~Moreira, I.~Amaral, I.~Domingues, A.~Cardoso, M.~J. Cardoso, and J.~S.
  Cardoso, ``I{NB}reast: toward a full-field digital mammographic database,''
  \emph{Academic Radiology}, vol.~19, no.~2, pp. 236--248, 2012.

\bibitem{cardoso2017mass}
J.~S. Cardoso, N.~Marques, N.~Dhungel, G.~Carneiro, and A.~P. Bradley, ``Mass
  segmentation in mammograms: A cross-sensor comparison of deep and tailored
  features,'' \emph{IEEE International Conference on Image Processing}, pp.
  1737--1741, 2017.

\bibitem{Heath2000digital}
M.~Heath, K.~Bowyer, D.~Kopans, R.~Moore, and W.~P. Kegelmeyer, ``The digital
  database for screening mammography,'' pp. 212--218, 2000.

\bibitem{Heath1998current}
M.~Heath, K.~Bowyer, D.~Kopans, W.~P. Kegelmeyer, R.~Moore, K.~Chang, and
  K.~MunishKumaran, ``Current status of the digital database for screening
  mammography,'' pp. 457--460, 1998.

\bibitem{Anmol2015}
\BIBentryALTinterwordspacing
A.~Sharma, ``D{DSM} utility,'' \emph{GitHub repository}, 2015. [Online].
  Available: \url{https://github.com/trane293/DDSMUtility}
\BIBentrySTDinterwordspacing

\bibitem{dhungel2015deep}
N.~Dhungel, G.~Carneiro, and A.~P. Bradley, ``Deep structured learning for mass
  segmentation from mammograms,'' \emph{International Conference on Image
  Processing}, pp. 2950--2954, 2015.

\bibitem{andreopoulos2008efficient}
A.~Andreopoulos and J.~K. Tsotsos, ``Efficient and generalizable statistical
  models of shape and appearance for analysis of cardiac {MRI},'' \emph{Medical
  Image Analysis}, vol.~12, no.~3, pp. 335--357, 2008.

\bibitem{glorot2010understanding}
X.~Glorot and Y.~Bengio, ``Understanding the difficulty of training deep
  feedforward neural networks,'' \emph{International Conference on Artificial
  Intelligence and Statistics}, pp. 249--256, 2010.

\bibitem{wachinger2017deepnat}
C.~Wachinger, M.~Reuter, and T.~Klein, ``Deep{NAT}: Deep convolutional neural
  network for segmenting neuroanatomy,'' \emph{NeuroImage}, vol. 170, pp.
  434--445, 2017.

\bibitem{shi2016learning}
Y.~Shi, Y.~Gao, S.~Liao, D.~Zhang, Y.~Gao, and D.~Shen, ``A learning-based {CT}
  prostate segmentation method via joint transductive feature selection and
  regression,'' \emph{Neurocomputing}, vol. 173, pp. 317--331, 2016.

\bibitem{dey2018diagnostic}
R.~Dey, Z.~Lu, and Y.~Hong, ``Diagnostic classification of lung nodules using
  {3D} neural networks,'' \emph{IEEE International Symposium on Biomedical
  Imaging}, pp. 774--778, 2018.

\bibitem{gecnn2016url}
\BIBentryALTinterwordspacing
``Group equivariant {CNN}.'' [Online]. Available:
  \url{https://github.com/adambielski/pytorch-gconv-experiments}
\BIBentrySTDinterwordspacing

\bibitem{li2019fully}
Z.~Li, Y.~Lou, Z.~Yan, S.~Alraref, J.~K. Min, L.~Axel, and D.~N. Metaxas,
  ``Fully automatic segmentation of short-axis cardiac mri using modified deep
  layer aggregation,'' in \emph{International Symposium on Biomedical Imaging},
  2019, pp. 793--797.

\bibitem{Duan2019Auto}
J.~Duan, J.~Schlemper, W.~Bai, and T.~J.~W. Dawes, ``Automatic {3D}
  bi-ventricular segmentation of cardiac images by a shape-refined multitask
  deep learning approach,'' \emph{IEEE Transactions on Medical Imaging},
  vol.~38, no.~9, pp. 2151--2164, 2019.

\bibitem{du2019card}
X.~Du, S.~Yin, R.~Tang, Y.~Zhang, and S.~Li, ``Cardiac-deepied: Automatic
  pixel-level deep segmentation for cardiac bi-ventricle using improved
  end-to-end encoder-decoder network,'' \emph{IEEE Journal of Translational
  Engineering in Health and Medicine}, vol.~7, pp. 1--10, 2019.

\bibitem{zotti2017gridnet}
C.~Zotti, Z.~Luo, O.~Humbert, A.~Lalande, and P.~Jodoin, ``Grid{Net} with
  automatic shape prior registration for automatic {MRI} cardiac
  segmentation,'' \emph{International Workshop on Statistical Atlases and
  Computational Models of the Heart}, pp. 73--81, 2018.

\bibitem{van2009python}
R.~G. Van and F.~Drake, ``Python 3 reference manual,'' \emph{Paramount (CA):
  CreateSpace}, 2009.

\bibitem{segcaps2018code}
\BIBentryALTinterwordspacing
``Seg{C}aps,'' \emph{GitHub}, 2018. [Online]. Available:
  \url{https://github.com/lalonderodney/SegCaps}
\BIBentrySTDinterwordspacing

\end{thebibliography}

\balance

\end{document}